\documentclass[iop]{emulateapj}

\newcommand{\HI}{\mathrm{H\,I}}

\newcommand{\GHI}{\Gamma_{\HI}}

\newcommand{\mfp}{\lambda_{\mathrm{mfp}}}

\newcommand{\lya}{Ly$\alpha$ }
\newcommand{\lyb}{Ly$\beta$ }

\newcommand{\teff}{\tau_\mathrm{eff}}
\newcommand{\chimp}{Mpc$/h$}
\newcommand{\xhi}{x_\mathrm{HI}}

\begin{document}
\title{Determining the Nature of Late Gunn-Peterson Troughs with Galaxy Surveys}
\author{Frederick B. Davies$^{1,2}$, George D. Becker$^3$, Steven R. Furlanetto$^4$}
\affil{$^1$Max-Planck-Institut f{\"u}r Astronomie, K{\"o}nigstuhl 17, D-69117 Heidelberg, Germany}
\affil{$^2$Department of Physics, University of California, Santa Barbara, CA 93106-9530, USA}
\affil{$^3$Department of Physics \& Astronomy, University of California, Riverside, CA 92521, USA}
\affil{$^4$Department of Physics \& Astronomy, University of California, Los Angeles, CA 90095-1547, USA}
\email{davies@physics.ucsb.edu}

\begin{abstract}
Recent observations have discovered long (up to $\sim 110$~Mpc$/h$), opaque Gunn-Peterson troughs in the $z \sim5.5$ Ly$\alpha$ forest, which are challenging to explain with conventional models of the post-reionization intergalactic medium. Here we demonstrate that observations of the galaxy populations in the vicinity of the deepest troughs can distinguish two competing models for these features: deep voids where the ionizing background is weak due to fluctuations in the mean free path of ionizing photons would show a deficit of galaxies, while residual temperature variations from extended, inhomogeneous reionization would show an overdensity of galaxies. We use large ($\sim550$ Mpc$/h$) semi-numerical simulations of these competing explanations to predict the galaxy populations in the largest of the known troughs at $z\sim5.7$. We quantify the strong correlation of Ly$\alpha$ effective optical depth and galaxy surface density in both models and estimate the degree to which realistic surveys can measure such a correlation. While a spectroscopic galaxy survey is ideal, we also show that a relatively inexpensive narrowband survey of Ly$\alpha$-emitting galaxies is $\sim90\%$ likely to distinguish between the competing models.
\end{abstract}

\section{Introduction}

The reionization of the Universe was the last major phase transition of baryons. The measured large-scale polarization of the cosmic microwave background (CMB) constrains the epoch of reionization to a characteristic redshift of $z\sim7$--$8$ \citep{Planck16b}, and observations of transmission in the \lya forest suggest that it has ended by $z\sim6$ \citep{Fan06,McGreer11,McGreer15}.

Above $z\sim5.5$, the \lya forest begins to exhibit large-scale ($\gtrsim 50$ \chimp) regions with transmission consistent with zero, commonly known as Gunn-Peterson (GP) troughs due to their similarity with expectations for a fully neutral IGM \citep{GP65}. However, saturated absorption in the \lya forest only requires volume-averaged hydrogen neutral fractions of $\xhi\sim10^{-4}$, so the presence of these troughs does not unambiguously indicate a neutral IGM \citep{Fan06}. GP troughs at $z\gtrsim5.5$ are not ubiquitous -- the opacity of the \lya forest, typically quantified by the effective optical depth $\teff\equiv-\ln{\langle F \rangle}$ (where $\langle F \rangle$ is the mean transmitted flux), measured on scales of $\sim50$ \chimp, varies considerably between different lines of sight \citep{Fan06}. 

Early studies suggested that these variations were consistent with expectations from the non-linear connection between the density field of the IGM and \lya forest $\teff$ \citep{Lidz06b}. However, more recent work by \citet[][henceforth B15]{Becker15} added new lines of sight to the original \citet{Fan06} sample and found that above $z\sim5.5$ the variations in $\teff$ are substantially stronger than the expectations from the density field alone, implying large fluctuations in the neutral hydrogen fraction ($\xhi$). Most notably, B15 discovered a $\sim110$ \chimp\ GP trough from $z\sim5.52$--$5.88$ ($\teff\geq7.4$ at $2\sigma$) in a deep spectrum of the quasar ULAS J0148+0600 ($z_\mathrm{Q}=5.98$). At these redshifts, the IGM appears to be predominantly highly ionized, and model-independent analysis of dark regions in the \lya and \lyb forests suggests that reionization is complete \citep{McGreer11,McGreer15}. Indeed, B15 found that the ULAS J0148+0600 trough has significant (albeit weak) transmission throughout the Ly$\beta$ forest, so it is unlikely that the trough as a whole represents a not-yet-reionized patch of neutral gas.

The opacity of the ionized IGM at $z\gtrsim5.5$ is closely linked to the neutral fraction of gas at low density ($\Delta\equiv\rho/\bar{\rho} < 1$, where $\bar{\rho}$ is the cosmic mean density; see, e.g., \citealt{OF05}). The equilibrium neutral fraction at fixed gas density is a function of two quantities: the photoionization rate of hydrogen $\GHI$ and the gas temperature $T$, where $\xhi\propto\GHI^{-1}T^{-0.7}$. Thus, stronger-than-expected (i.e. not directly due to the density field) variations in $\teff$ are either due to large fluctuations in the ionizing background or in the temperature of the IGM. In particular, the environment of the giant GP trough in B15 either has a weak ionizing background or is much colder than the typical IGM.

B15 showed that simple models for fluctuations in the ionizing radiation field due to the clustering of galaxies and a spatially-uniform mean free path of ionizing photons ($\mfp$) were unable to reproduce the giant GP trough in ULAS J0148+0600 while remaining in agreement with the $\teff$ in the typical \lya forest at $z\sim5.7$. \citet[][henceforth DF16]{DF16} expanded upon the radiation field modeling of B15 by introducing spatial variations in $\mfp$. Fundamentally, $\mfp$ represents absorption of ionizing photons by residual neutral hydrogen in the IGM, so a stronger (weaker) ionizing background should result in a longer (shorter) $\mfp$ due to the change in $\xhi$. DF16 applied the analytic model of \citet{McQuinn11} to self-consistently compute ionizing background and $\mfp$ fluctuations in a large semi-numerical simulation of the IGM. The enhanced fluctuations in the ionizing background due to $\mfp$ fluctuations were able to reproduce the observed distribution of $\teff$, provided the average $\mfp$ was smaller by a factor of two than an extrapolation from $\mfp$ measurements\footnote{This discrepancy can be accounted for by a likely bias in the $\mfp$ measured along quasar sightlines at $z\ga5$ (see \citealt{D'Aloisio17} for details).} at $z\sim2.5$--$5.2$ \citep{Worseck14,D'Aloisio17}.  Giant GP troughs in this model result from large-scale \emph{voids} which are distant from sources of ionizing photons, and thus whose local $\mfp$ is very short, strongly suppressing the ionizing background \citep[see also][]{Crociani11}. 

The process of reionization deposits a substantial amount of thermal energy into the IGM from excess energy in newly-photoionized electrons, leading to (initially) isothermal post-reionization gas temperatures of $\sim1$--$4\times10^4$ K depending on the nature of ionizing photon sources \citep{MR94,AH99,McQuinn12,Davies16}. The gas then cools, predominantly through inverse Compton scattering off of the CMB and adiabatic cooling from the expansion of the Universe, such that by $z\lesssim5$ the thermal state of the IGM is largely independent of the details of the reionization process (\citealt{Theuns02,HH03}, although see \citealt{Trac08,Cen09}). However, the typical cooling timescale is of order the Hubble time, so if reionization ended as late as $z\sim6$ there may be substantial residual heat present in the IGM at $z\sim5.5$. Additionally, because the reionization process is inhomogeneous on large ($\ga10$ Mpc) scales \citep{Furlanetto04}, this residual heat could vary greatly between different regions of the Universe. \citet{D'Aloisio15} investigated the possibility of large-scale residual temperature fluctuations from the inhomogeneous reionization process as the source of large variations in $\teff$ through the temperature dependence of the hydrogen recombination rate. \citet{D'Aloisio15} found that temperature fluctuations from a hot ($\Delta T = 3\times10^4$ K) and extended ($z\sim13$--$6$) epoch of reionization were sufficient to source the large observed scatter in $\teff$. In their model, large regions with high $\teff$ are \emph{overdense} regions which reionized early, and thus have had more time to cool than the typical IGM.

An alternative source of strong fluctuations in the ionizing background (and thus, $\teff$) could instead be a sparsity of ionizing sources relative to the mean free path \citep{Zuo92,MW03}, e.g. if the ionizing background is dominated by luminous quasars \citep{Chardin15}. Even this model, however, appears to require coupled fluctuations of the mean free path and ionizing background to be consistent with observations \citep{Chardin17}. Constraints from the thermal state of the IGM at lower redshift and helium reionization suggest that a large contribution to the ionizing photon budget from quasars at $z\ga5$ is unlikely \citep{D'Aloisio17a}. For simplicity, we do not explore it further in this work. Finally, we note that \citet{Gnedin17} examined the sensitivity of reionization models to other \lya forest statistics, including the distribution of dark gap lengths and the characteristics of transmission peaks. In this work we will focus primarily on the $\teff$ measurements from \citet{Fan06} and B15.

Thus, we can make two opposing predictions for the environment of the giant GP trough in ULAS J0148+0600: either the region is \emph{underdense} and has a much weaker ionizing background than average (DF16, \citealt{D'Aloisio17}), or the region is \emph{overdense} and is much cooler than average \citep{D'Aloisio15}. In this work, we study the implications of these two models on the cross-correlation between GP troughs and their local galaxy populations, with particular focus on the giant GP trough in ULAS J0148+0600. In \S~2, we describe our semi-numerical modeling of ionizing background and temperature fluctuations in a $\sim550$ \chimp\ cosmological volume and their resulting distributions of $\teff$ at $z\sim5.7$. In \S~3, we predict the distribution of UV-bright galaxies, abundance-matched to dark matter halos in our semi-numerical simulation, in large GP trough environments in the two scenarios. In \S~4, we make similar predictions for Ly$\alpha$-emitting galaxies by applying the conditional \lya equivalent width model of \citet{DW12}. Finally, in \S~5 we conclude with a summary and prospects for other methods.

In this work we assume a $\Lambda$CDM cosmology with $h=0.7$, $\Omega_m=0.3$, $\Omega_\Lambda=0.7$, $\Omega_b=0.048$ and $\sigma_8=0.82$. Distance units are comoving unless specified otherwise.

\section{Semi-numerical Modeling of $\teff$ Fluctuations}

In this work, we used the semi-numerical code {\small DEXM} \citep{MF07} to produce a model for the density field of the IGM at $z=5.7$. We generated a $4096^3$ realization of cosmological initial conditions (ICs) in a volume $546$ Mpc$/h$ on a side. We then quasi-linearly evolved the ICs to $z=5.7$ using the Zel'dovich approximation \citep{Zel'dovich70} onto a coarser $2048^3$ grid including the oversampling prescription of DF16 to reduce shot noise in the evolved density field. The left panel of Figure~\ref{fig:densityhalos} shows a 50 Mpc$/h$-thick slice of the density field.

\subsection{Fluctuations in the ionizing background due to $\mfp$ fluctuations}\label{sec:uvbcalc}

\begin{figure*}
\begin{center}
\resizebox{16cm}{!}{\includegraphics[trim={2.0em 2.5em 1.0em 6.0em},clip]{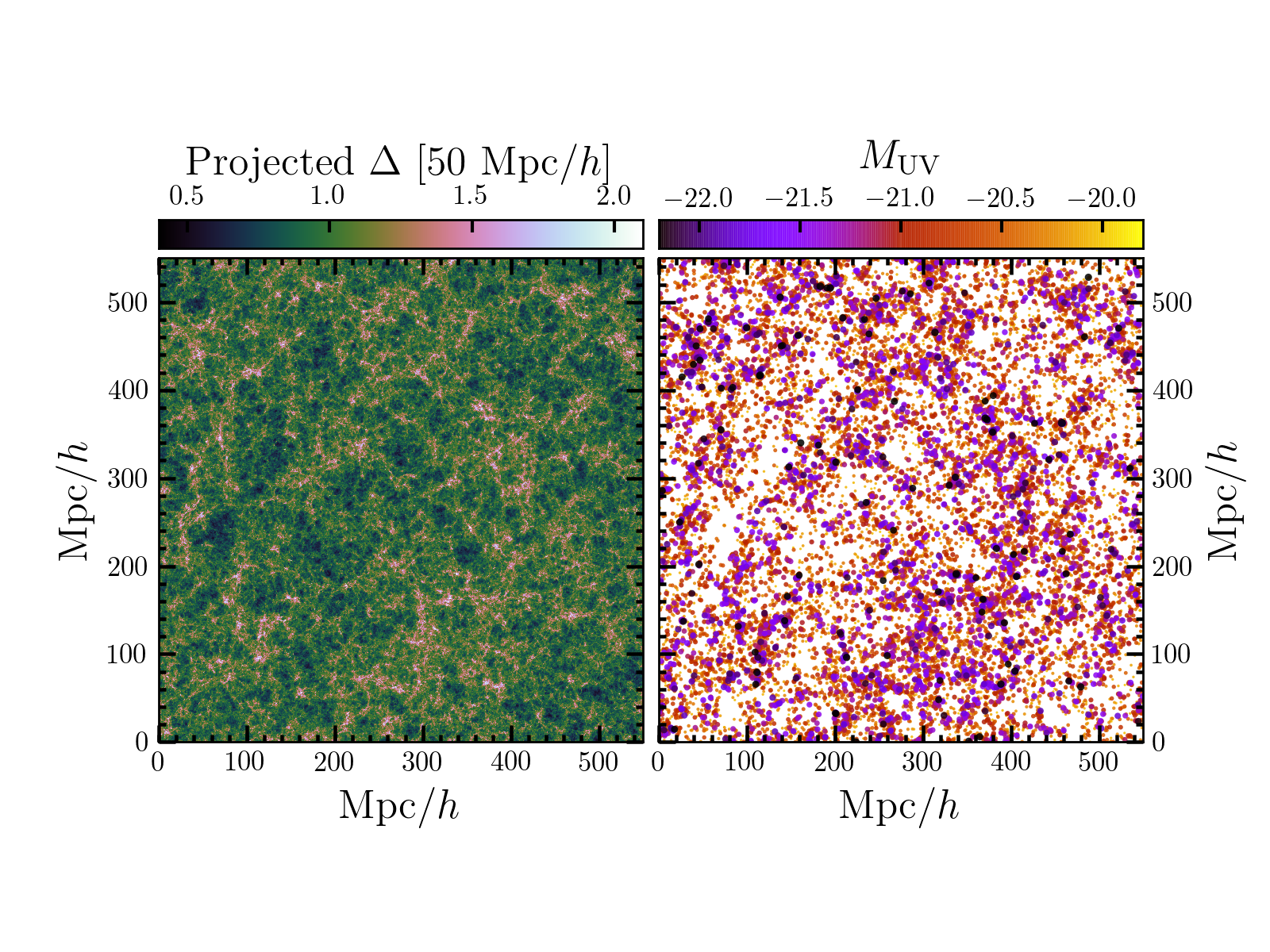}}\\
\end{center}
\vskip -3.75em
\caption{Left: Semi-numerical density field averaged over 50 Mpc$/h$ perpendicular to the page. Right: Distribution of dark matter halos within the same volume as the left panel, color- and size-coded by $M_\mathrm{UV}$ determined via abundance matching. Only halos with $M_\mathrm{UV}\la-20$ are shown, but our model includes a complete distribution of halos down to $M_{\rm UV,max}\approx-12.9$.}
\label{fig:densityhalos}
\end{figure*}

To compute large-scale fluctuations in the ionizing background in our cosmological volume we employed the method of DF16, which we briefly summarize here. Dark matter halos with masses greater than $M_\mathrm{min}=2\times10^9$ M$_\odot$ were located in the volume using an excursion-set-like procedure on the ICs linearly evolved to $z=5.7$ \citep{MF07}, and then their positions were updated using the same Zel'dovich displacement field that was used to evolve the density field. Rest-frame UV luminosities (i.e. galaxies) were assigned to each halo by abundance matching\footnote{The star formation efficiency of dark matter halos implied by this approach appears to be reasonable at high redshift \citep{Mirocha17}.} (e.g. \citealt{VO04}) to the measured UV luminosity function\footnote{In detail, we interpolate between the $z\sim5.9$ and $z\sim4.9$ luminosity functions to estimate the luminosity function at $z=5.7$.} of \citet{Bouwens15}, with our minimum halo mass corresponding to $M_{\rm UV,max}\approx-12.9$. The distribution of $M_\mathrm{UV}<-20$ galaxies in a 50 Mpc$/h$-thick slice of the simulation is shown in the right panel of Figure~\ref{fig:densityhalos}, demonstrating their highly clustered nature. We further assume that the ionizing luminosity of each galaxy is proportional to its UV luminosity, and leave the ratio of ionizing-to-UV luminosity as a free (but constant) parameter. The spatially-variable opacity of the IGM, parameterized by $\mfp$, was assumed to follow the relation $\mfp\propto\GHI^{2/3}\Delta^{-1}$, motivated by analytic arguments in \citet[][see the discussion in DF16 for details]{McQuinn11}. We adopt $\mfp(\GHI=\langle\GHI\rangle,\Delta=1)=15$ Mpc, which DF16 found was required to explain the distribution in $\teff$ measured by B15 at $z\sim5.6$.

\begin{figure*}
\begin{center}
\resizebox{16cm}{!}{\includegraphics[trim={2.0em 2.5em 1.0em 6.0em},clip]{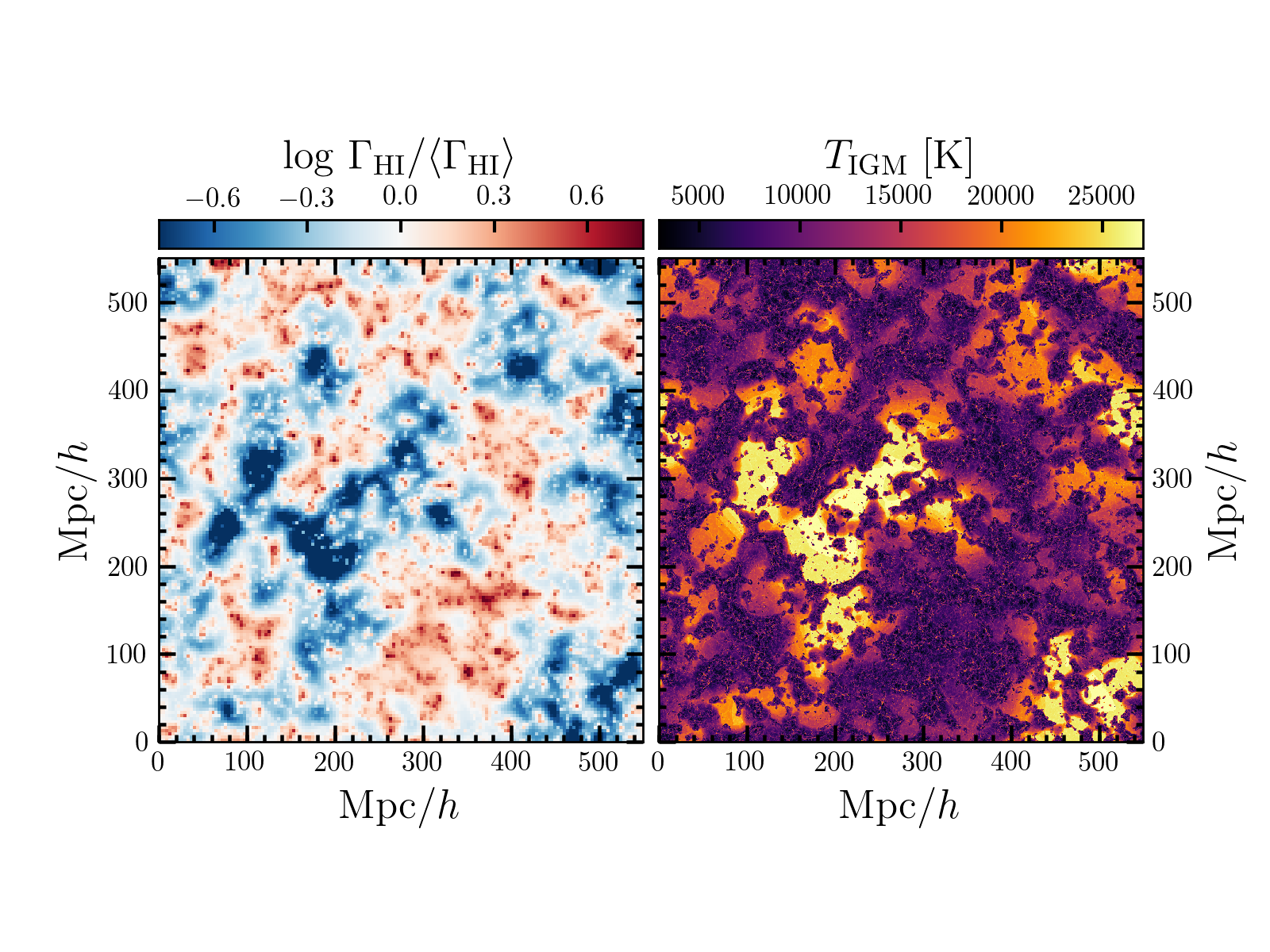}}\\
\end{center}
\vskip -3.75em
\caption{Left: 5 Mpc-thick (1 cell) slice through the fluctuating ionizing background model computed in Section~\ref{sec:uvbcalc}. Features are apparent on scales up to $\sim100$ Mpc$/h$, as required by the ULAS J0148+0600 trough. Right: 400 kpc-thick (1 cell) slice through the residual temperature fluctuations model, computed via numerical integration of equation~(\ref{eqn:tempeq}) with an initial heat input of $\Delta T=3\times10^{4}$ K at the reionization redshift (Figure~\ref{fig:zreion}).}
\label{fig:uvbtemp}
\end{figure*}

The ionizing background was then iteratively computed on a $156^3$ grid, corresponding to the same 5 Mpc resolution as DF16. We show a slice of the $\GHI$ field in the left panel of Figure~\ref{fig:uvbtemp}, demonstrating correlated fluctuations on scales as large as $\sim100$ \chimp. In this model, overdense regions have an enhanced $\GHI$, while voids have a very weak $\GHI$. The inclusion of $\mfp$ fluctuations is critical to the magnitude of this effect, especially in weakening the radiation field in void environments.

\begin{figure}
\begin{center}
\resizebox{8cm}{!}{\includegraphics[trim={8.5em 0 11em 0},clip]{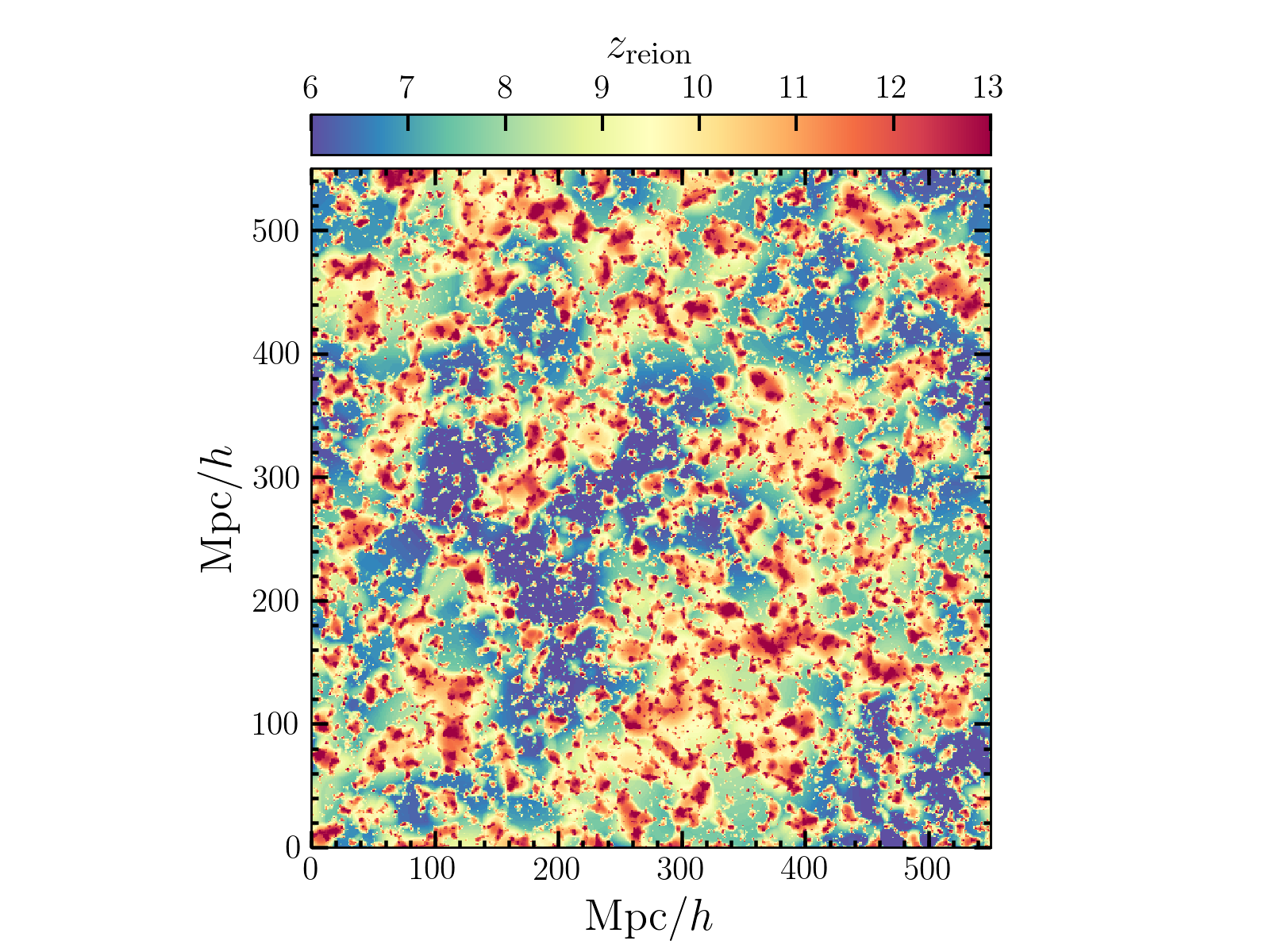}}\\
\end{center}
\caption{$\sim800$ kpc-thick (1 cell) slice through the reionization redshift field used to compute IGM temperatures. Overdense regions are reionized first, while underdense regions are reionized later, following the generic predictions of inside-out reionization in the excursion set model.}
\label{fig:zreion}
\end{figure}

\subsection{Fluctuations in IGM temperature due to inhomogeneous and extended reionization}\label{sec:tempcalc}

We model post-reionization fluctuations in the temperature of the IGM using a procedure similar to \citet{D'Aloisio15} but in a semi-numerical fashion more analogous to the method of \citet{LM14}. We compute the reionization redshift field ($z_\mathrm{reion}$) in our cosmological volume on a 1024$^3$ grid by computing excursion set reionization maps in {\small DEXM} from $z=5.8$--$13.0$ in steps of $\Delta z=0.1$. For computational efficiency, we computed the reionization maps using settings in {\small DEXM} that mimic the code {\small 21CMFAST} \citep{Mesinger11} -- that is, the fraction of material in collapsed objects was computed analytically from the density field itself rather than the halo field, and we adopt the central-pixel-flagging algorithm instead of the slower bubble-painting algorithm (see \citealt{Zahn11} for a discussion of the relative accuracy of these methods). The minimum halo mass for producing ionizing photons was fixed to $M_\mathrm{min}=2\times10^9 M_\odot$, for consistency with the halos in the ionizing background model, and the ionizing efficiency $\zeta$ was tuned as a function of redshift to produce an ionization history very similar to that in \citet{D'Aloisio15}: reionization begins at $z\sim13$, and then $\xhi(z)$ evolves roughly linearly with redshift until reionization is largely complete by $z\sim6$. A slice of the resulting reionization redshift field is shown in Figure~\ref{fig:zreion} (compare to the left panel of fig. 3 in \citealt{D'Aloisio15}).

To compute the residual temperature field at $z=5.7$, we employ a simple temperature evolution model similar to that described in \citet{UptonSanderbeck16}. We assume that, at $z=z_\mathrm{reion}$, each cell in our cosmological volume is heated to $3\times10^4$ K.\footnote{This is at the high end of plausible values for the IGM temperature after passage of ionization fronts during reionization by sources with a cutoff at the \ion{He}{2} ionizing edge \citep{McQuinn12}, regulated by collisional excitation cooling as the front passes a gas parcel \citep{Cantalupo08,Davies16}. For our assumed reionization history, such intense heating is required to reproduce the strength of $z\ga5.5$ \lya forest fluctuations \citep{D'Aloisio15}.} The temperature evolution of gas in the IGM is then evolved as \citep{HG97,FO08,UptonSanderbeck16}
\begin{equation}\label{eqn:tempeq}
\frac{dT}{dt} = -2HT + \frac{2T}{3\Delta}\frac{d\Delta}{dt} - \frac{T}{n_\mathrm{tot}}\frac{dn_\mathrm{tot}}{dt} + \frac{2}{3k_Bn_\mathrm{tot}}\frac{dQ}{dt},
\end{equation}
where the first term describes adiabatic cooling due to Hubble expansion, the second term describes adiabatic heating or cooling due to density evolution, and the third term describes heating or cooling due to energy equipartition between different species of particles where $n_\mathrm{tot}$ is the total number density of baryons (atoms, ions, \& electrons). The final term folds all additional heating and cooling processes into $dQ/dt$; in this work, we include photoionization heating due to the \ion{H}{1} and \ion{He}{1} ionizing backgrounds, for which we assume a power law $J_\nu\propto\nu^{-1.5}$ from 1 to 4 Ryd and zero intensity above 4 Ryd consistent with the pre-\ion{He}{2} reionization Universe, recombination cooling, collisional ionization cooling, collisional excitation cooling from \ion{H}{1} line emission, free-free emission, and inverse Compton cooling off of the CMB.\footnote{We use recombination rates and recombination cooling rates from \citet{HG97}, collisional ionization cooling rates from \citet{Theuns98}, and collisional excitation cooling rates from \citet{Cen92}.} For simplicity, we follow \citet{UptonSanderbeck16} and assume a Zel'dovich pancake model for the density evolution of gas elements, $\Delta(a) = [1 - \lambda G(a)]^{-1}$, where $a$ is the scale factor, $G(a)$ is the growth factor, and $\lambda$ is a constant normalized by reproducing $\Delta$ at $z=5.7$. In principle we could compute the Zel'dovich approximated density field for a finely sampled range of redshifts to more consistently compute the density evolution of each cell, but differences in the resulting temperatures would likely be small \citep{MUS16}.

We show how the temperature-density relation of gas depends on reionization redshift in Figure~\ref{fig:tempevol}. Gas at relatively high overdensities ($\Delta>10$) quickly cools to an equilibrium state, but gas at low density retains substantial excess heat for $\Delta z>1$ \citep{Trac08,FO09,D'Aloisio15}. We then apply this temperature evolution model to each cell in the $2048^3$ semi-numerical cosmological density field through a precomputed interpolation table of $T(z=5.7)$ as a function of reionization redshift $z_\mathrm{reion}$ and $\Delta(z=5.7)$. The resulting median temperature-density relation is shown by the orange curve in Figure~\ref{fig:tempevol}, which is similar to the post-reionization state of the IGM found in other works \citep{Trac08,FO09,LM14}. In the right panel of Figure~\ref{fig:uvbtemp} we show a $\sim400$ kpc (1 cell) slice of the resulting IGM temperature field. Large voids where reionization finished late are hot and nearly isothermal with $T\sim2.5\times10^4$ K, while overdense regions which reionized early have relaxed to a power-law temperature-density relation with $T\sim7500$ K at the mean density.

\subsection{Distribution of \lya forest effective optical depth}

\begin{figure}
\begin{center}
\resizebox{8cm}{!}{\includegraphics{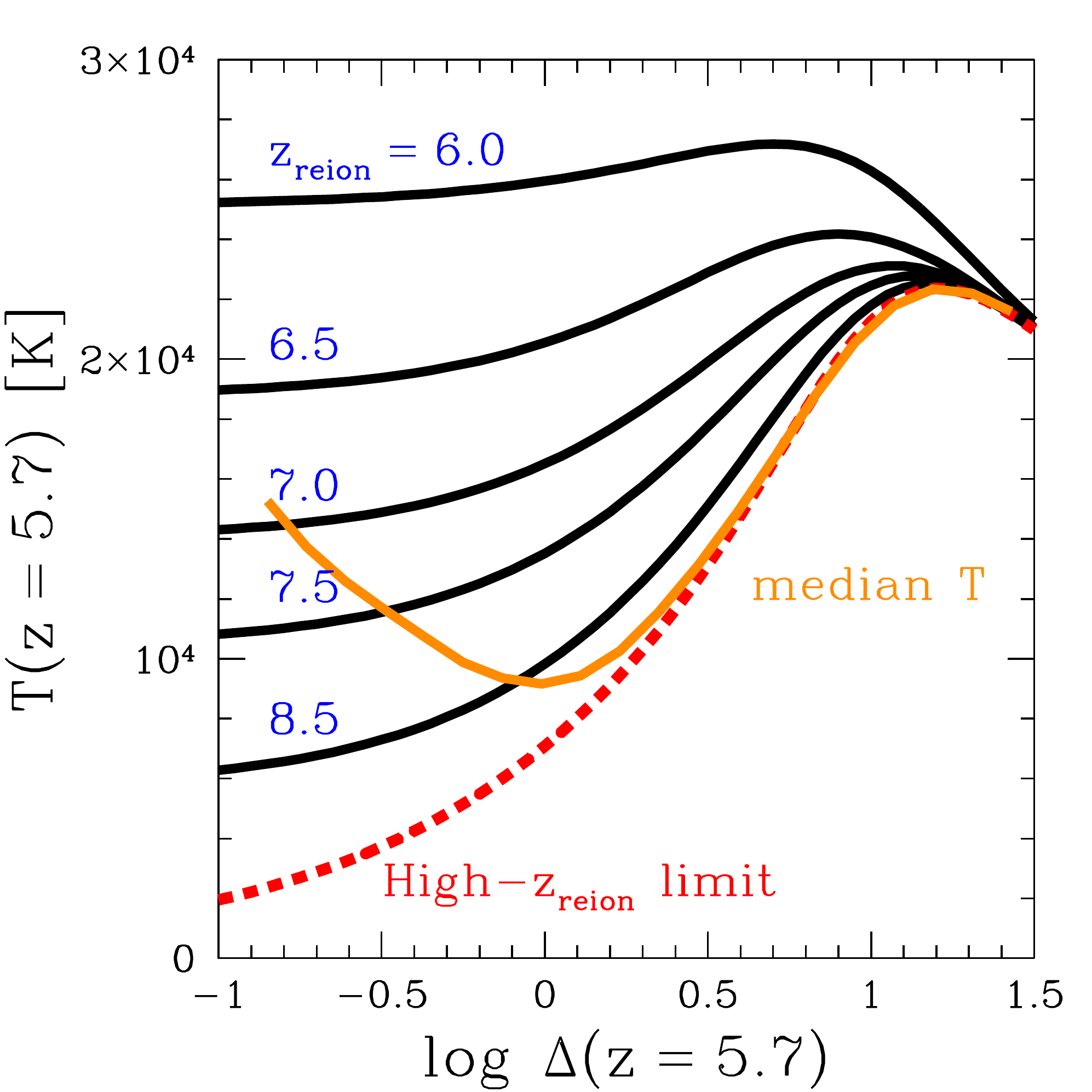}}\\
\end{center}
\caption{Evolution of the IGM temperature-density relation following equation~\ref{eqn:tempeq} for different reionization redshifts $z_\mathrm{reion}$ (black curves). Reionization is assumed to impulsively heat the IGM to $3\times10^4$ K, then the IGM cools due to inverse Compton cooling off of CMB photons, adiabatic cooling from the expansion of the Universe, recombination cooling, and collisional excitation cooling. The red dashed curve shows the IGM temperature-density relation in the limit when reionization occurred very early. The orange curve shows the median temperature-density relation of our semi-numerical simulation, computed in bins of dlog$\Delta=0.12$.}
\label{fig:tempevol}
\end{figure}

We compute $\teff$ for many random sightlines through the semi-numerical density field using the fluctuating Gunn-Peterson approximation (FGPA; e.g. \citealt{Weinberg97}),
\begin{eqnarray}\label{eqn:taugp}
\tau_\mathrm{GP} &\approx& 36.1\kappa\,\Delta^{2}\left( \frac{T}{7500\,\mathrm{K}} \right)^{-0.724} \nonumber \\
&\times& \left( \frac{\GHI}{10^{-12.5}\,\mathrm{s}^{-1}} \right)^{-1}  \left( \frac{1+z}{6.7} \right)^{4.5},
\end{eqnarray}
where $\kappa$ is a re-normalization constant that we tune to match the median observed $\teff$ in the Ly$\alpha$ forest on large scales ($\approx 3.5$, B15). In the fluctuating $\GHI$ model, we assume that the gas in the IGM follows a power-law temperature density relation $T=T_0 \Delta^{\gamma-1}$ \citep{HG97} with $T_0=7500$ K and $\gamma=1.5$, consistent with the early reionization heating curve in Figure~\ref{fig:tempevol}. In the residual temperature fluctuations model, we assume a uniform ionizing background with $\GHI=3\times10^{-13}$ s$^{-1}$. We additionally compute skewers assuming constant $\GHI=3\times10^{-13}$ s$^{-1}$ and $T=(7500$~K)~$\Delta^{0.5}$ to represent the expected $\teff$ fluctuations from the density field alone -- this is similar to the typical state of hydrodynamical simulations at $z=5.7$ which do not include impulsive models for heating during reionization (e.g. \citealt{Puchwein15,Onorbe17}). While this simple model for the \lya forest is unable to produce realistic transmission profiles for individual spectra, the distribution of $\teff$ measured on large scales and how the distribution changes due to $\GHI$ fluctuations is similar to hydrodynamical simulations \citep{D'Aloisio17}.

The required value of the constant $\kappa$ mostly reflects our neglect of peculiar velocities and thermal broadening in the calculation of $\teff$ from equation~\ref{eqn:taugp}, but also more subtle effects such as the limited spatial resolution of our semi-numerical density field. To match the observed median $\teff\sim3.5$ at $z\sim5.7$, 
we require $\kappa=$ 0.14, 0.31 for the ionizing background and temperature fluctuations models, respectively. The large difference between the two $\kappa$ values reflects the fact that, in the temperature fluctuations model, low density gas responsible for \lya transmission \citep{OF05} has a much higher typical temperature.

\begin{figure}
\begin{center}
\resizebox{8cm}{!}{\includegraphics{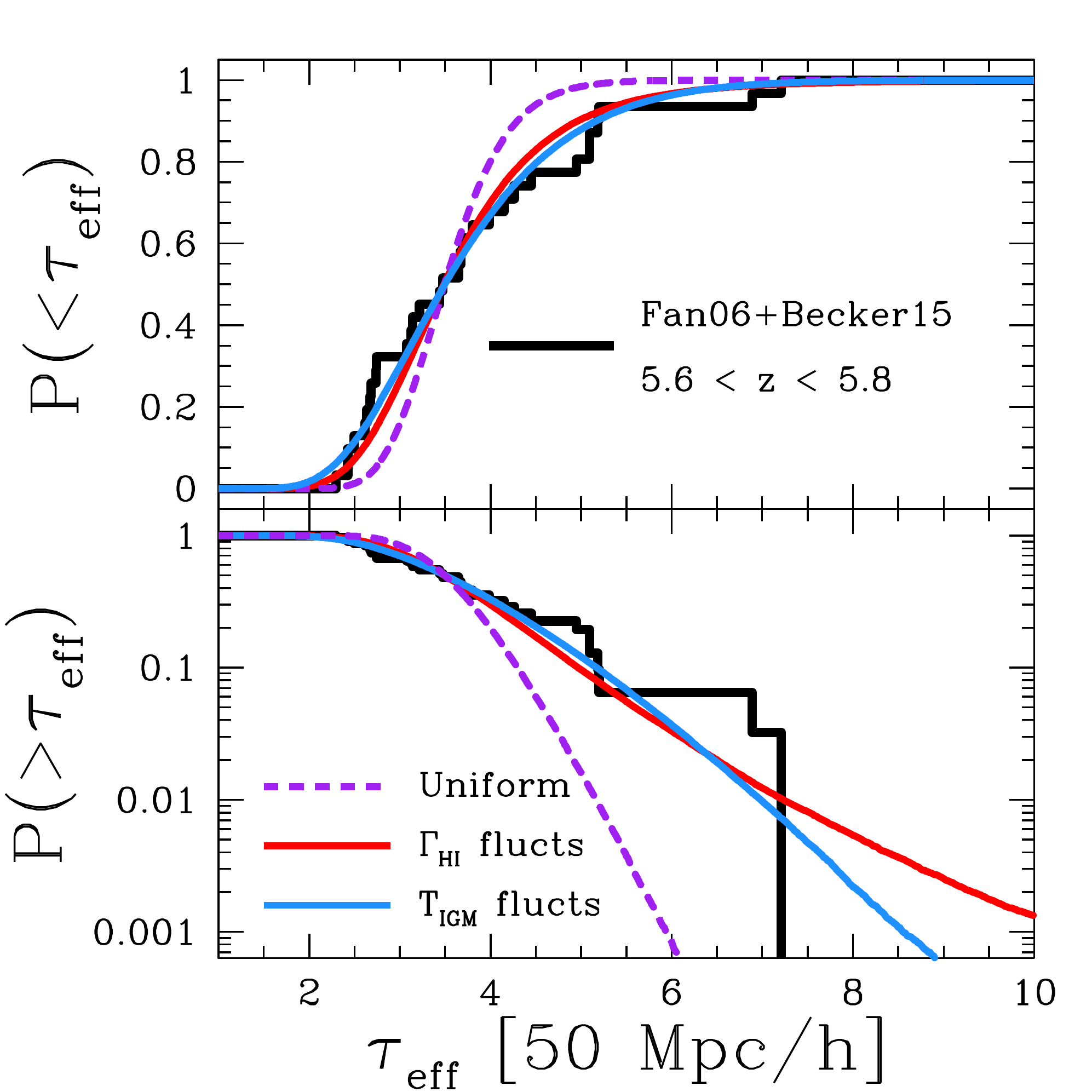}}\\
\end{center}
\caption{Top: Cumulative distribution of $\teff$ in the uniform model (dashed purple), including ionizing background fluctuations (red), and including temperature fluctuations (blue). The black curve shows the cumulative distribution of $\teff$ measurements in Fan et al. (2006) and Becker et al. (2015) with central redshifts between $5.6 < z < 5.8$, including $2\sigma$ lower limits.}
\label{fig:taudist}
\end{figure}

In the top panel of Figure~\ref{fig:taudist} we show the resulting cumulative distribution of $\teff$ on 50 Mpc$/h$ scales for the uniform, fluctuating $\GHI$, and fluctuating $T_{\rm IGM}$ models. Both fluctuating models broaden the $\teff$ distribution relative to the uniform model, and more closely match a compilation of 31 measurements of $\teff$ with central redshifts between $5.6 < z < 5.8$ from \citet{Fan06} and B15.\footnote{We note that a handful of the data points in this cumulative distribution actually represent $2\sigma$ \emph{lower limits} to $\teff$, most importantly the two measurements at $\teff\sim7$ in the giant GP trough of ULAS J0148+0600 from B15.} In the lower panel of Figure~\ref{fig:taudist} we focus on the high-$\teff$ tail of the distribution, where the differences between the models themselves and with the data are the most pronounced. The fluctuating ionizing background model shows an extended tail at $\teff\ga6$ while the temperature fluctuations model falls more rapidly, but at the current sensitivity of the limits of the deepest GP trough in ULAS J0148+0600 (and given the current number of measurements), the two models are largely indistinguishable. This similarity is in qualitative agreement with \citet{D'Aloisio17}, who found that the modifications to the \lya forest power spectrum on large-scales due to fluctuations in $\GHI$ and $T_{\rm IGM}$ were roughly equivalent.

Clearly, the distribution of $\teff$ in the \lya forest is unable to distinguish between the two scenarios for fluctuations that we have considered. One potential avenue for breaking this degeneracy is the difference in the physical nature of GP troughs implied by the models, as we discuss in the following section.

\begin{figure*}
\begin{center}
\resizebox{16cm}{!}{\includegraphics[trim={2.5em 6.5em 2.0em 8.0em},clip]{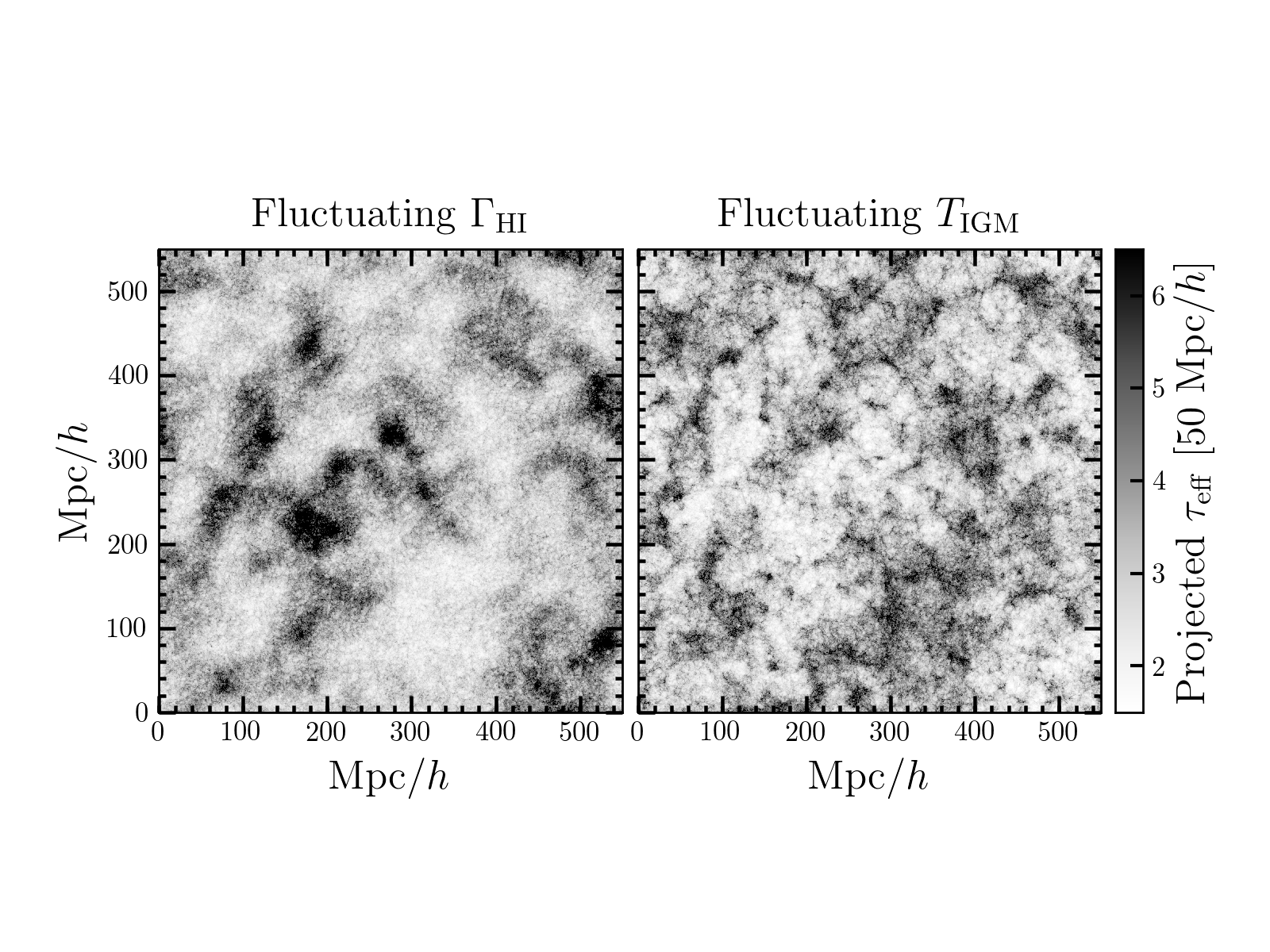}}\\
\end{center}
\caption{50 Mpc$/h$-projected \lya forest $\teff$ for the fluctuating ionizing background (left) and residual temperature fluctuations (right) models, centered on the slice shown in Figure~\ref{fig:uvbtemp}. The regions corresponding to low and high $\teff$ correspond to high and low density regions, respectively, for the fluctuating ionizing background model, but the opposite is true for the residual temperature fluctuations model.}
\label{fig:taumaps}
\end{figure*}

\begin{figure*}
\begin{center}
\resizebox{16cm}{!}{\includegraphics[trim={2.5em 6.5em 1.0em 6.0em},clip]{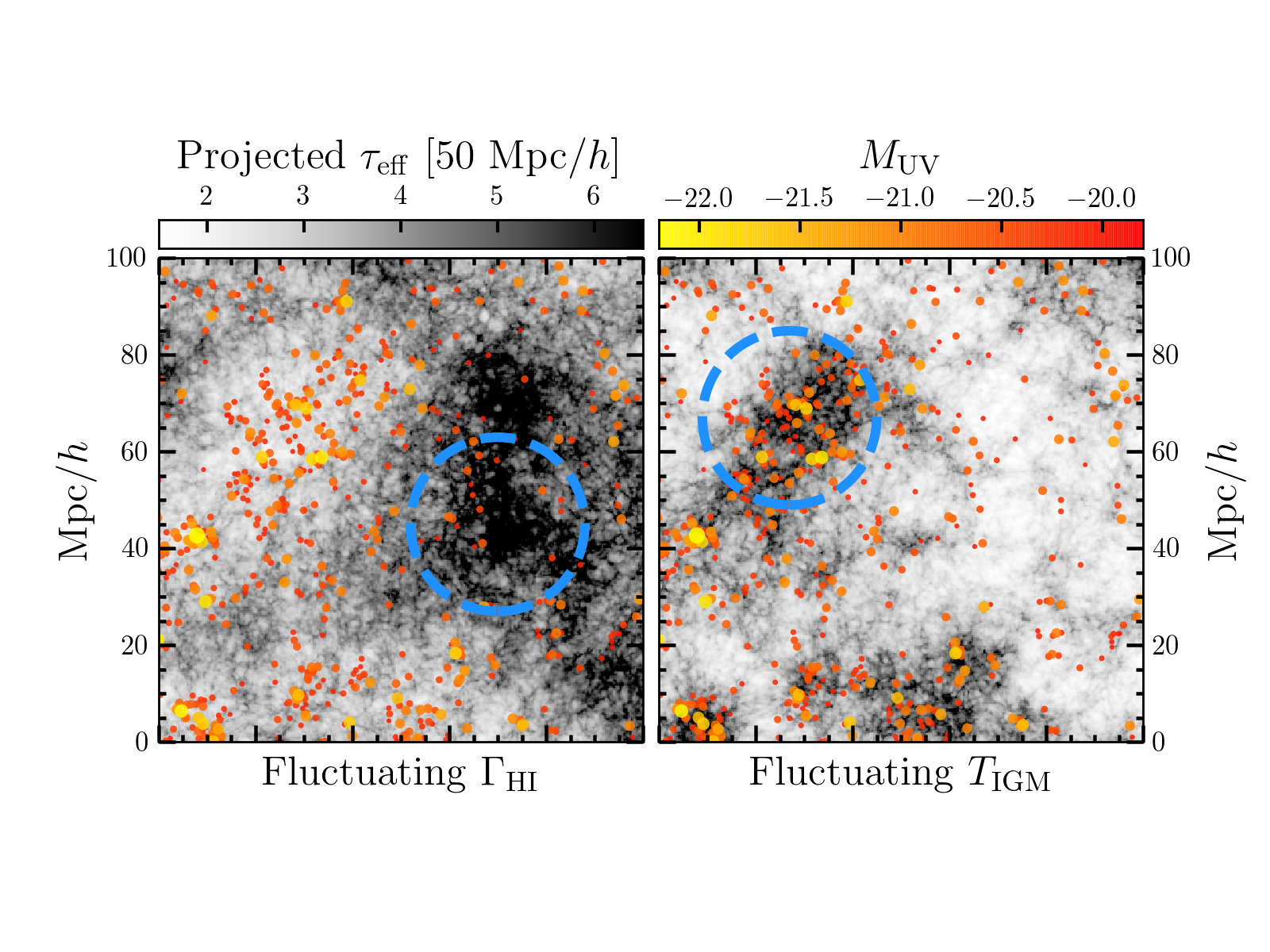}}\\
\end{center}
\caption{Zoomed-in portion (100 Mpc$/h$ on a side) of the 50 Mpc$/h$-projected \lya forest $\teff$ for the fluctuating ionizing background (left) and residual temperature fluctuations (right) models with overlaid $M_{\rm UV}<-20$ galaxies (dots) color- and size-coded by absolute magnitude. The corresponding region in Figure~\ref{fig:taumaps} is centered on (500, 370) Mpc$/h$. The dashed circles correspond to 10 arcmin radius patches of sky centered on a high $\teff$ sightline in each model.}
\label{fig:taumaps_zoom}
\end{figure*}

\section{Relationship between \lya forest opacity and galaxy~populations}

The two models for post-reionization $\teff$ fluctuations that we consider, ionizing background fluctuations and residual temperature fluctuations, make opposite predictions for the nature of GP troughs. If GP troughs are due to ionizing background fluctuations (as in DF16), they arise in large-scale \emph{voids} that are far away from sources of ionizing photons and exhibit a short mean free path. If GP troughs are instead the result of residual temperature fluctuations (as in \citealt{D'Aloisio15}), they arise in large-scale \emph{overdense regions} that reionized at early times and have had time to cool. In Figure~\ref{fig:taumaps}, we show maps of the 50 Mpc$/h$-scale $\teff$ centered on the slice shown in Figure~\ref{fig:uvbtemp} -- compare to the projected density field in the left panel of Figure~\ref{fig:densityhalos}.

While the 100 Mpc$/h$ over- or under-density associated with these regions should be numerically small due to the homogeneity of the Universe on such large scales, galaxies at high redshift are highly biased tracers of the density field \citep{Overzier09}. Thus, in principle, the local galaxy population of GP troughs could favor one model over the other. Our semi-numerical simulation is well-suited to predicting the strength of the correlation between \lya forest opacity and the co-spatial galaxy population on these large scales. As discussed in Section~\ref{sec:uvbcalc}, we have abundance matched the halos in our simulation to the observed UV luminosity function of $z\sim6$ galaxies \citep{Bouwens15}. In Figure~\ref{fig:taumaps_zoom}, we show a close-up on a ($100$ Mpc$/h$)$^2$ section of the $\teff$ field with overlaid $M_{\rm UV}<-20$ galaxies, demonstrating a strong spatial relationship between high-$\teff$ regions and bright galaxies which reverses between the $\GHI$ and $T_{\rm IGM}$ fluctuations models. To investigate the statistical nature of this relationship, we have computed 1,000,000 randomly oriented lines of sight through both of our fiducial \lya opacity simulations, described in the previous section, and counted the number of galaxies co-spatial with the \lya forest region (i.e. within the redshift range of the measured $\teff$) as a function of transverse distance from the sightline and $M_\mathrm{UV}$. 

\begin{figure}

\begin{center}
\resizebox{8.5cm}{!}{\includegraphics[trim={2em 11em 1.5em 0},clip]{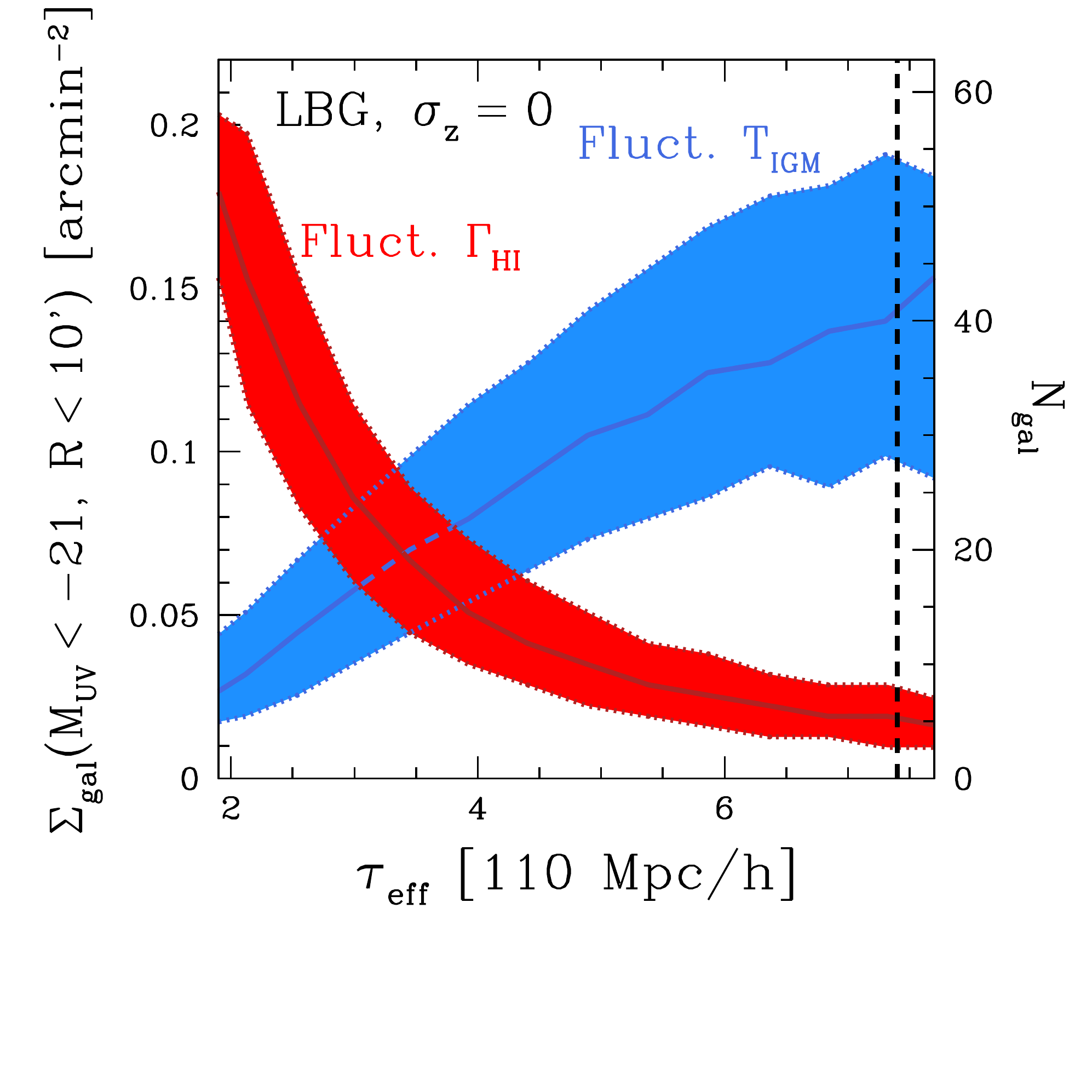}}\\
\end{center}
\caption{Relationship between the number of $M_\mathrm{UV}<-21$ galaxies within a transverse sky separation of 10 arcmin and 110 Mpc$/h$ \lya forest effective optical depth for the fluctuating ionizing background model (red) and residual temperature fluctuations model (blue). The vertical dashed line corresponds to the $\teff$ lower limit of the J0148+0600 GP trough. At high (and low) $\teff$, the distributions of galaxy counts are highly distinct.}
\label{fig:taugal_bb}
\end{figure}

\begin{figure}
\begin{center}
\resizebox{8.5cm}{!}{\includegraphics[trim={2em 11em 1.5em 0},clip]{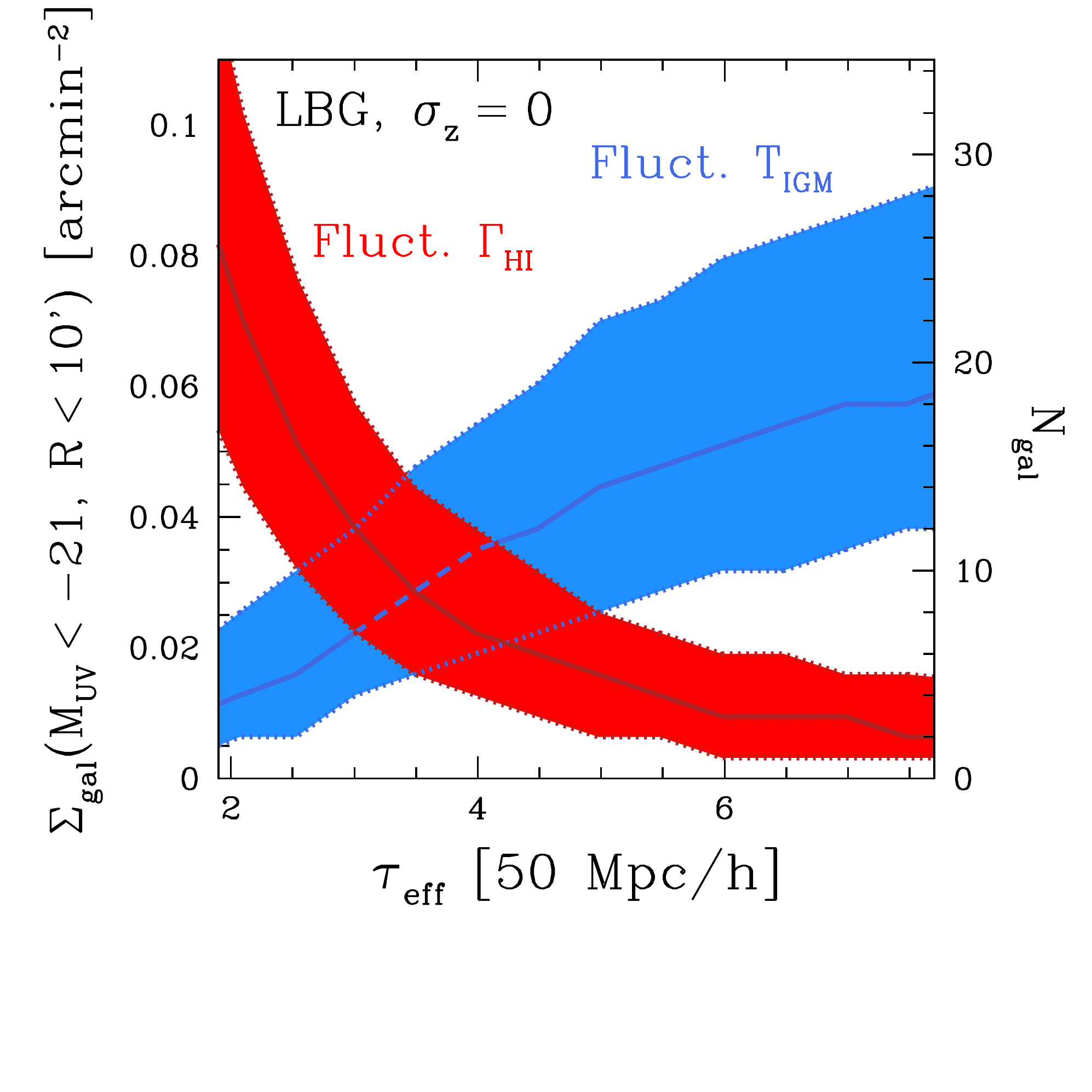}}\\
\end{center}
\caption{Same as Figure~\ref{fig:taugal_bb} but for $\teff$ and galaxy counts measured on 50 Mpc$/h$ scales.}
\label{fig:taugal_bb_50}
\end{figure}

In Figure~\ref{fig:taugal_bb}, we show the correlation between the number of bright ($M_\mathrm{UV}<-21$) galaxies within a transverse separation of 10 arcmin ($\sim18$ Mpc$/h$) and $\teff$ on 110 Mpc$/h$ scales for our two fluctuations models. The solid lines represent the median number of galaxies at each $\teff$, while the shaded region shows the 16--84th percentiles of the galaxy count distribution. The two models exhibit the trends predicted above -- high $\teff$ corresponds to relatively low or high density regions, and the strong difference in galaxy counts at $\teff\ga6$ reflects the large bias of their host halos. In Figure~\ref{fig:taugal_bb_50} we show a similar test but for $\teff$ and galaxy counts measured on 50 Mpc$/h$ scales. The trends are weaker because the smaller spatial scale probes less extreme outliers at fixed $\teff$ and the scatter in galaxy counts is higher.

\begin{figure}

\begin{center}
\resizebox{8.5cm}{!}{\includegraphics[trim={0 0 0 0},clip]{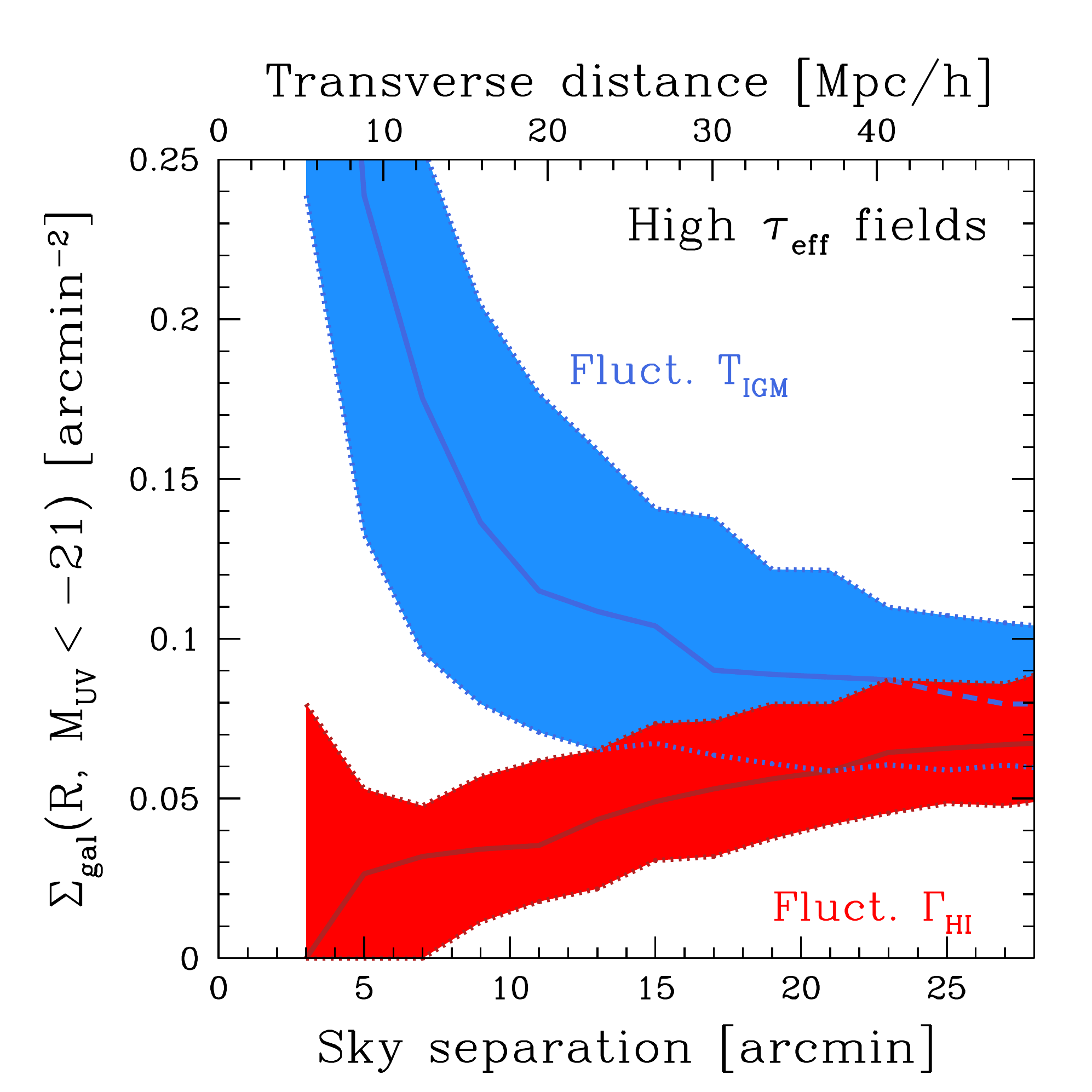}}\\
\end{center}
\caption{Radial profile of $M_\mathrm{UV}<-21$ galaxy density for $\teff>6.5$ sightlines in the fluctuating ionizing background model (red) and residual temperature fluctuations model (blue). The solid and dashed curves show the median and 16--84th percentiles of the galaxy count distributions, respectively. The difference in galaxy surface density persists to very large scales, with the difference in the median galaxy surface density persisting out to $\ga30$ arcmin.}
\label{fig:taugalr_bb}

\end{figure}

We explore the radial dependence of these correlations in Figure~\ref{fig:taugalr_bb} for 110 Mpc$/h$ sightlines with $\teff>6.5$, showing the surface density of $M_\mathrm{UV}<-21$ galaxies as a function of separation on the sky. The correlation between galaxy counts and $\teff$ decreases with distance from the \lya forest sightline, but is still non-negligible out to separations of $\ga20$ arcmin ($\ga 36$ Mpc$/h$).

\begin{figure}

\begin{center}
\resizebox{8.5cm}{!}{\includegraphics[trim={2em 11em 1.5em 0},clip]{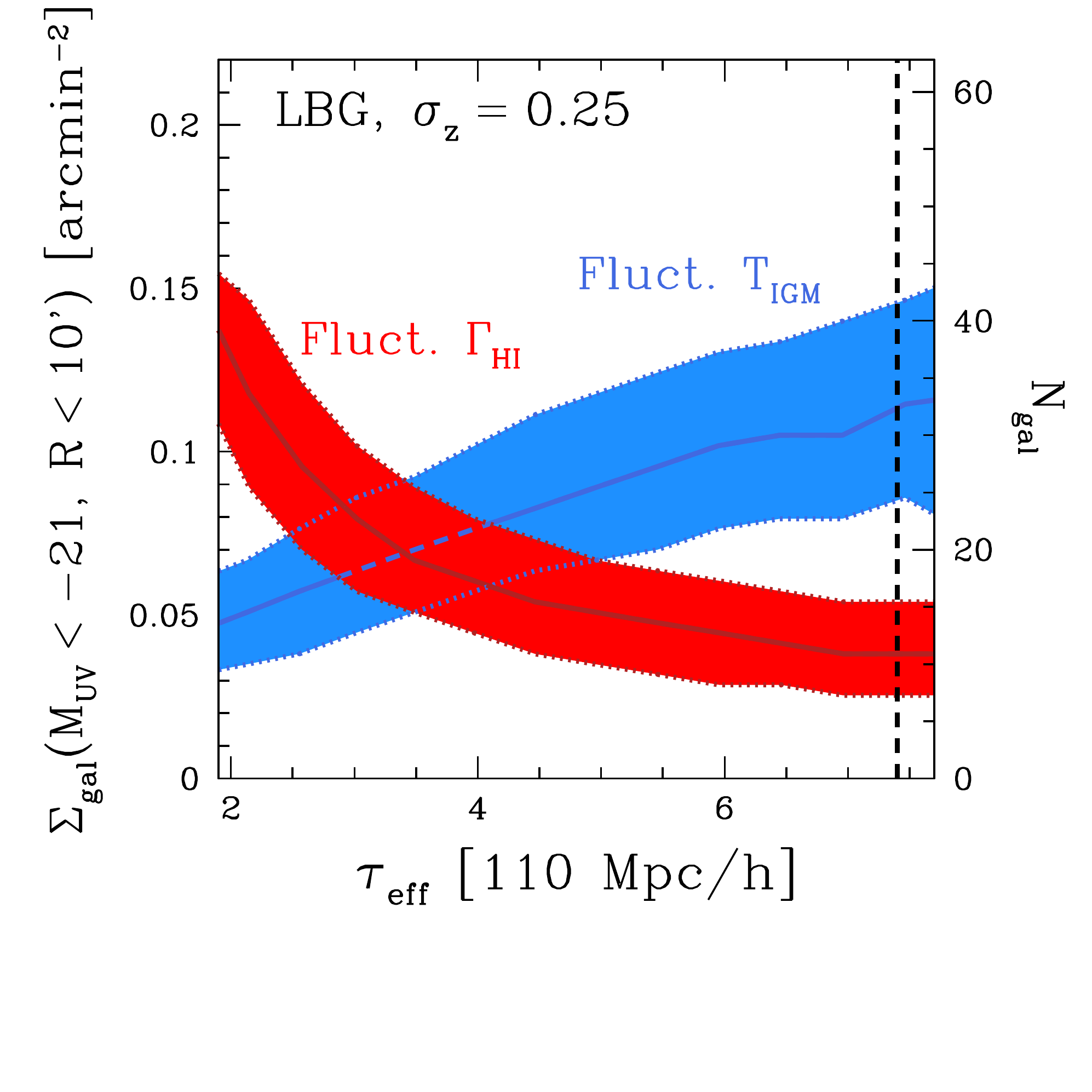}}\\
\end{center}
\caption{Same as Figure~\ref{fig:taugal_bb}, but assuming line-of-sight scatter of galaxies due to uncertain photometric redshifts ($\sigma_z=0.25$). Redshift uncertainty increases the effective region over which galaxies are selected -- outside the region where $\teff$ is measured -- thus reducing the strength of the correlation.}
\label{fig:taugal_bb_photz}

\end{figure}

In practice, counting $z\sim5.7$ galaxies in a large area of the sky presents an observational challenge. While surveys searching for bright high-redshift galaxies using broadband color-selection are relatively inexpensive, the precision of photometric redshifts is limited to scales comparable to the window over which $\teff$ is measured. In Figure~\ref{fig:taugal_bb_photz} we show the effect of a photometric redshift uncertainty $\sigma_z=0.25$ on the correlation between galaxies and $\teff$. Even for this relatively modest photometric redshift uncertainty ($\sigma_z/(1+z)\sim4\%$) the difference between the two models as compared to Figure~\ref{fig:taugal_bb} is greatly reduced due to galaxies scattering into (and out of) the target redshift range where $\teff$ was measured. Thus, expensive spectroscopic followup would be required to accurately count galaxies associated with the target region of the \lya forest. 

\section{Predictions for \lya emitters in the ULAS~J0148+0600 GP trough environment}

\begin{figure*}
\begin{center}
\resizebox{16cm}{!}{\includegraphics[trim={2.5em 6.5em 1.0em 6.0em},clip]{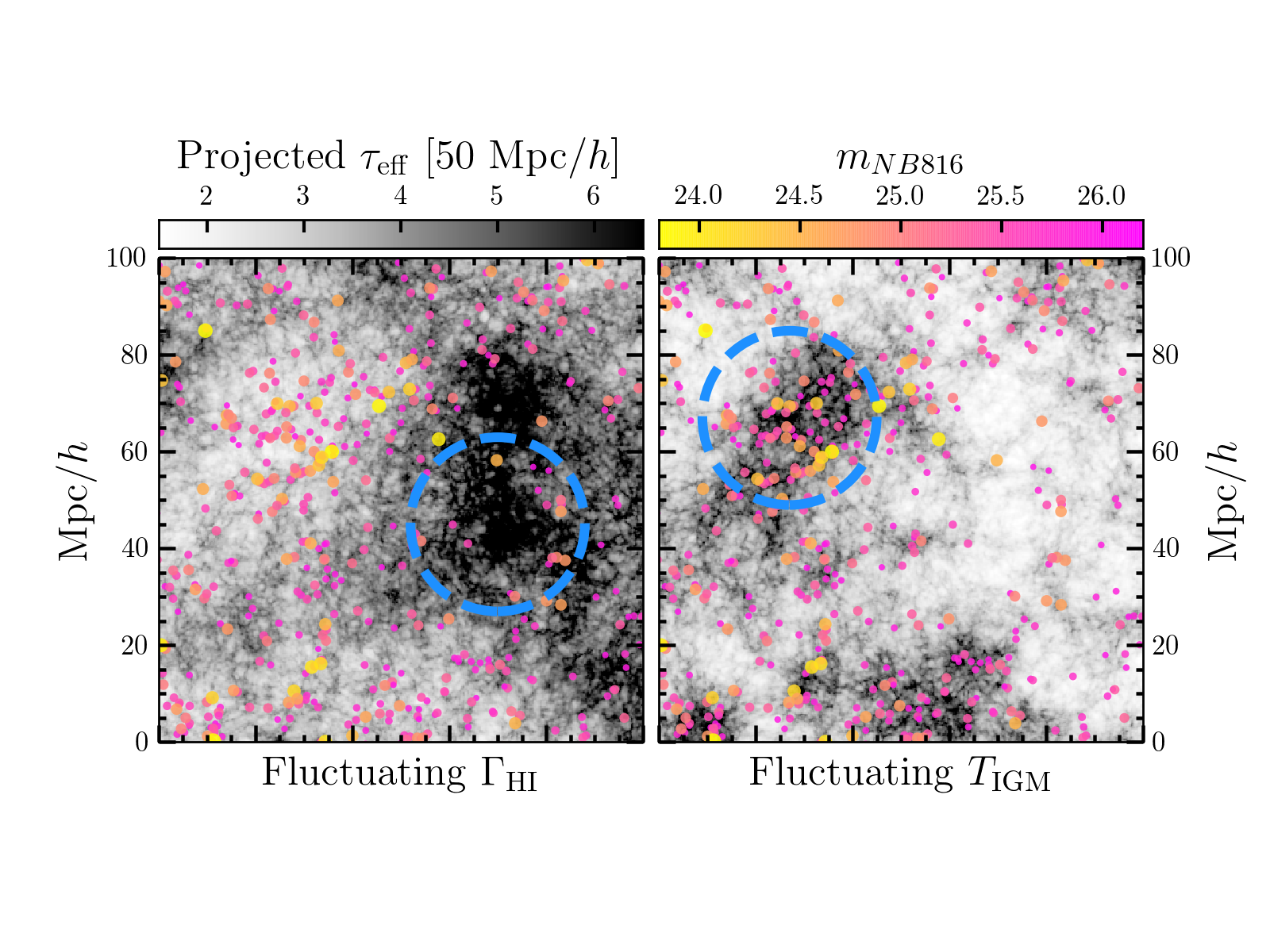}}\\
\end{center}
\caption{Similar to Figure~\ref{fig:taumaps_zoom} but with overlaid $m_{\rm NB816}<26$ LAEs (dots), color- and size-coded by apparent $NB816$ UV magnitude. Note that the LAEs shown here often do not directly correspond to galaxies shown in Figure~\ref{fig:taumaps_zoom} because their UV magnitude can be fainter than $M_{\rm UV}=-20$ and thus they were not shown.}
\label{fig:taumaps_zoom_lae}
\end{figure*}

\begin{figure}
\begin{center}
\resizebox{8.5cm}{!}{\includegraphics[trim={2.5em 0 1.0em 0},clip]{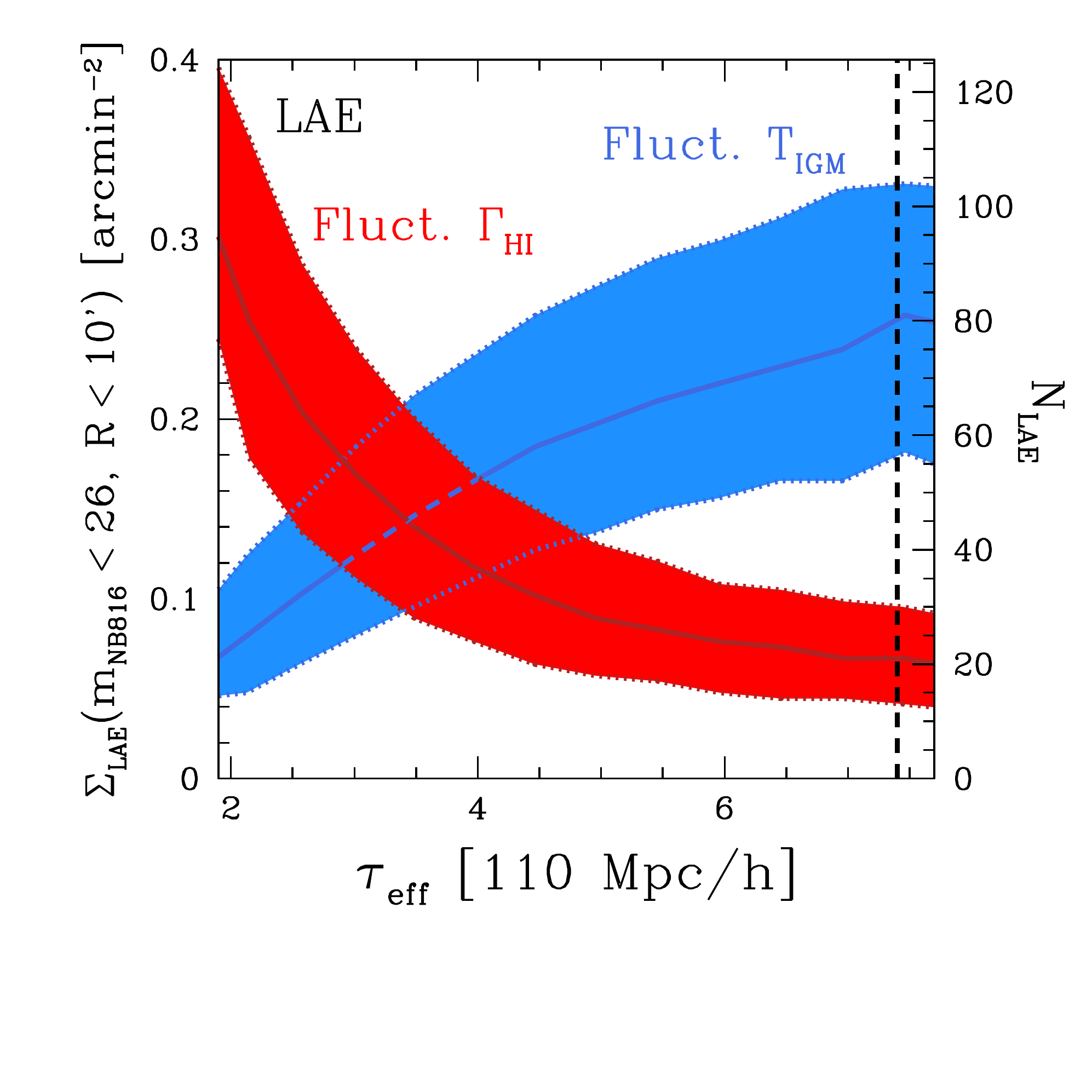}}\\
\end{center}
\vskip -5em
\caption{Relationship between the number of $m_{NB816}<26$ LAEs within a transverse sky separation of 10 arcmin and 110 Mpc$/h$ \lya forest effective optical depth for the fluctuating ionizing background model (red) and residual temperature fluctuations model (blue). The vertical dashed line corresponds to the $\teff$ lower limit of the J0148+0600 GP trough.}
\label{fig:taugal_nb}
\end{figure}

\begin{figure}
\begin{center}
\resizebox{8.5cm}{!}{\includegraphics[trim={2.5em 0 1.0em 0},clip]{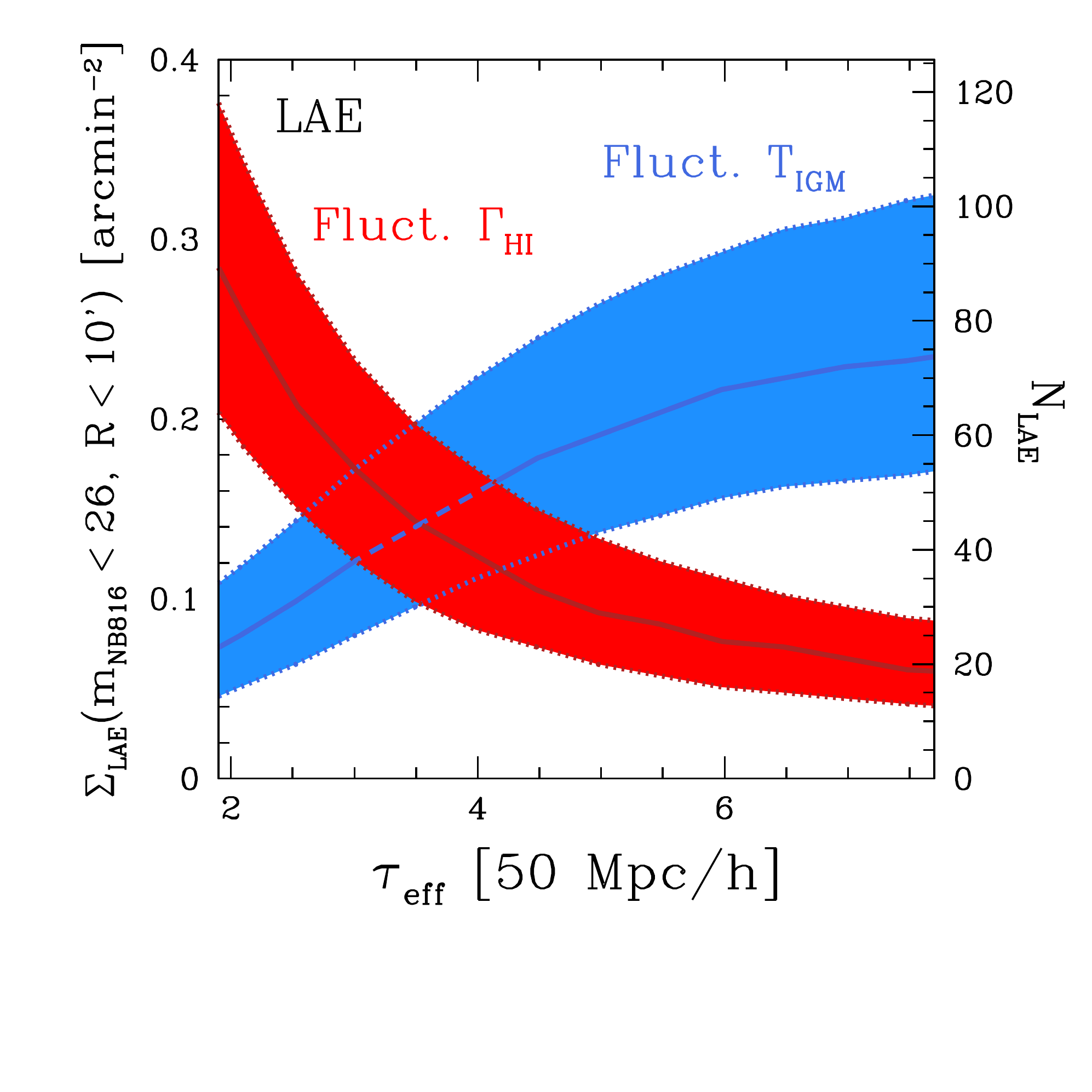}}\\
\end{center}
\vskip -5em
\caption{Same as Figure~\ref{fig:taugal_nb} but for $\teff$ measured on 50 Mpc$/h$ scales.}
\label{fig:taugal_nb_50}
\end{figure}

A relatively cheap alternative to spectroscopic galaxy surveys is narrowband color-selection of Ly$\alpha$-emitting galaxies (also known as \lya emitters, or LAEs). Fortuitously, \lya emission redshifted to $z\sim5.7$ corresponds to a peak in atmospheric transmission at $\sim8150$\AA, and thus narrowband filters capable of selecting LAEs at $z\sim5.7\pm0.05$ are readily available on several wide-field imaging instruments on large telescopes. In this section, we focus on predictions for an LAE survey with Hyper Suprime-Cam (HSC; \citealt{Miyazaki12}) on the 8.2-meter Subaru Telescope, which has a narrowband filter ($NB816$) well-suited to this task. 

To estimate the ability of LAEs to constrain the physical nature of the giant trough, we must associate the dark matter halos in our semi-numerical simulation to \lya line luminosities. We have decided to follow a simple empirical approach using the relation between $M_\mathrm{UV}$ and the rest-frame equivalent width (REW) of \lya from \citet[][henceforth DW12]{DW12}. In brief, the \lya REW of each galaxy is drawn from an exponential distribution,
\begin{equation}
P(\mathrm{REW}|M_\mathrm{UV}) \propto \exp{\left[\frac{-\mathrm{REW}}{\mathrm{REW}_c(M_\mathrm{UV})}\right]},
\end{equation}
where the scale length is a function of $M_\mathrm{UV}$: REW$_c = (22$ \AA$) + (6$ \AA$)\times(M_\mathrm{UV}+21.9)$. The distribution includes galaxies with negative REW (i.e. they show \lya in absorption), and is defined to be zero below REW$_\mathrm{min}=-20$ \AA\ for $M_\mathrm{UV}<-21.5$, REW$_\mathrm{min}=-20+6(M_\mathrm{UV}+21.5)^2$ \AA\ for $-19.0<M_\mathrm{UV}<-21.5$, and REW$_\mathrm{min}=-5$ \AA\ for $M_\mathrm{UV}>-19$. DW12 showed that this distribution was a good fit to the measured distribution of spectroscopically measured \lya REW for Lyman-break galaxies (LBGs) at $z=3$--$7$ \citep{Shapley03,Stark10}.

We select LAEs from the galaxies in our simulation as follows. The \lya forest is assumed to cover a redshift range $5.52 < z < 5.88$, and we assigned a redshift $z_\mathrm{gal}$ to each associated galaxy based on its location along the line of sight. We compute the $i$-band and $NB816$ magnitudes for each galaxy assuming a power-law UV continuum $F_\lambda\propto\lambda^{-2}$ plus the EW of its \lya line at $z = z_\mathrm{gal}$ and evolving IGM attenuation redward of \lya from the $\teff(z)$ measured by \citet{Fan06}. We then select galaxies as LAEs if $m_{NB816}<26$ and $i-NB816<1.2$, similar to the sensitivity and criterion of \citet{Ouchi08}, where the limiting narrowband magnitude is achievable with less than one night of exposure time with HSC. Direct application of the distribution from DW12 to the measured galaxy UV luminosity function, however, results in $\sim2$ times as many LAEs as actually detected by narrowband surveys \citep{Ouchi08}. DW12 found that a simple ``correction factor" $F=0.44$ to the LAE luminosity function computed from the distribution of REWs resulted in a very good fit to observations. We follow this correction in spirit by randomly throwing out 56\% of our simulated galaxies that would be selected as LAEs so that we reproduce the observed density of LAEs on the sky. While this is certainly an ad hoc correction, we confirm that our simulated LAE population matches the \lya luminosity function from \citet{Ouchi08}.

Figure~\ref{fig:taumaps_zoom_lae} shows that the relationship between $\teff$ and the underlying galaxy population (Figure~\ref{fig:taumaps_zoom}) is preserved when LAEs are selected. In Figure~\ref{fig:taugal_nb}, we show the analog of Figure~\ref{fig:taugal_bb} for LAEs within 10 arcmin of the \lya forest sightline. The ``signal" in LAE counts tends to be weaker than for the ideal case of a clean spectroscopic galaxy sample for two reasons: first, LAEs tend to lie in less massive (less biased) halos, and second, LAEs are only selected in a narrow $\sim30$ Mpc$/h$ window at the center of the sightline where the \lya line falls into the $NB816$ filter. The latter fact means that probing giant GP troughs much larger than the size of the \lya forest region may not matter, and indeed, in Figure~\ref{fig:taugal_nb_50} we show that the comparison between LAE counts and $\teff$ on 50 Mpc$/h$ scales is nearly identical. Despite the weaker distinction in LAE counts at the tails of the $\teff$ distribution compared to the spectroscopically-complete LBG case, the difference between our two models is quite stark, and should be detectable.

\begin{figure}
\begin{center}
\resizebox{8.5cm}{!}{\includegraphics{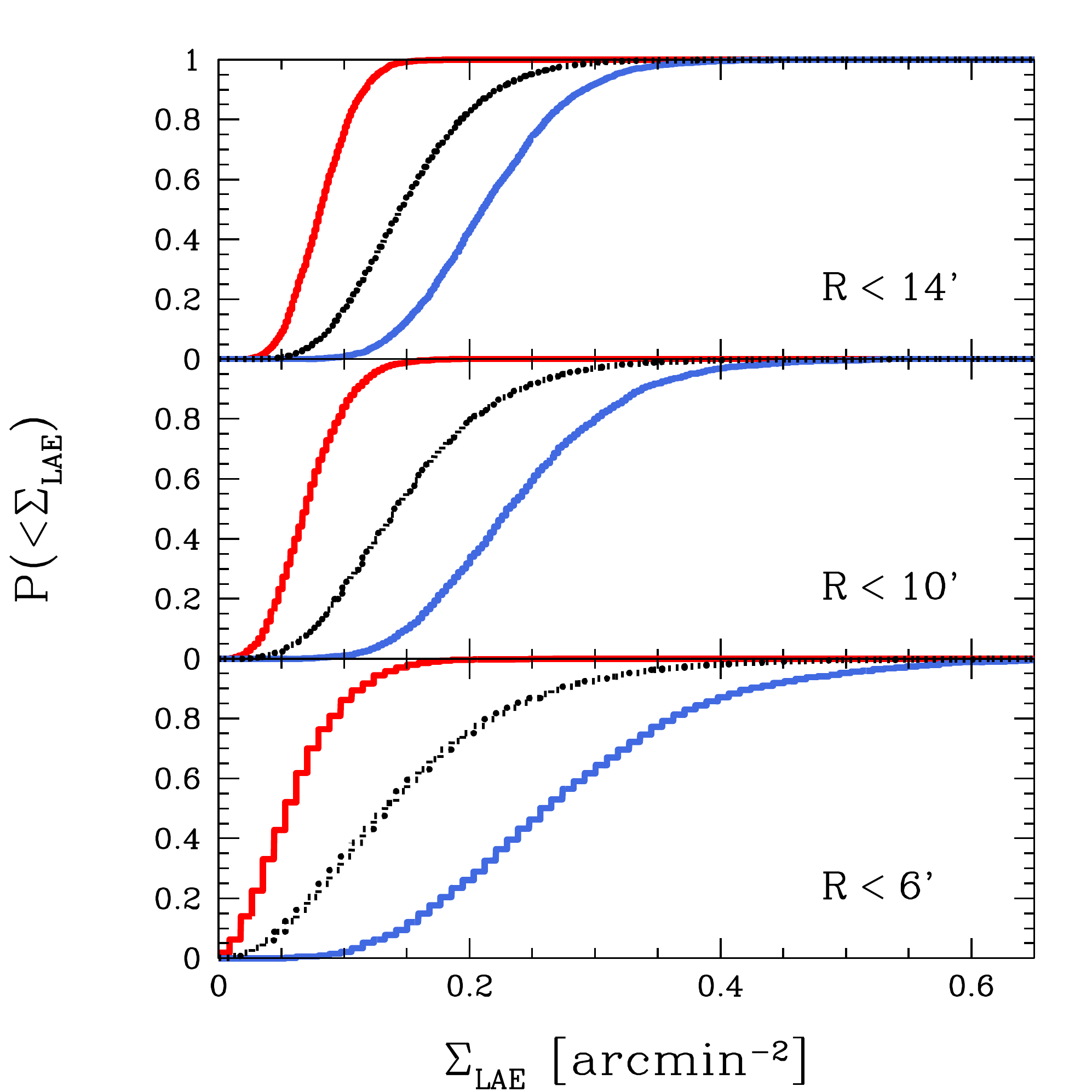}}\\
\end{center}
\caption{Cumulative distributions of $\Sigma_{\rm LAE}$ for the fluctuating $\GHI$ (red) and $T_{\rm IGM}$ (blue) models in high $\teff$ fields. The top, middle, and bottom panels show target radii of 14, 10, and 6 arcmin, respectively. The dotted curves show the distribution for all $\teff$, i.e. the expected distribution of random fields.}
\label{fig:lae_cumu}
\end{figure}

We show the cumulative distributions (CDF) of mock $\Sigma_{\rm LAE}$ measurements for the two models in Figure~\ref{fig:lae_cumu}, demonstrating that the distributions are moderately distinct. To quantify the ability of one measurement to distinguish between the two models, and to find the optimally sized region for comparison, we ask the question: if Model A is true, how likely is it that we rule out Model B at $95\%$ confidence? To estimate this, we use the CDFs of $\Sigma_{\rm LAE}$ (e.g. Figure~\ref{fig:lae_cumu}) from both models. Specifically, we compute CDF($\Sigma_{\rm LAE}\,|$ fluct. $\GHI$) with $\Sigma_{\rm LAE}$ corresponding to CDF($\Sigma_{\rm LAE}\,|$ fluct. $T_{\rm IGM}$) = 0.05, and CDF($\Sigma_{\rm LAE}\,|$ fluct. $T_{\rm IGM}$) with $\Sigma_{\rm LAE}$ corresponding to CDF($\Sigma_{\rm LAE}\,|$ fluct. $\GHI$) = 0.95. 

\begin{figure}

\begin{center}
\resizebox{8.5cm}{!}{\includegraphics{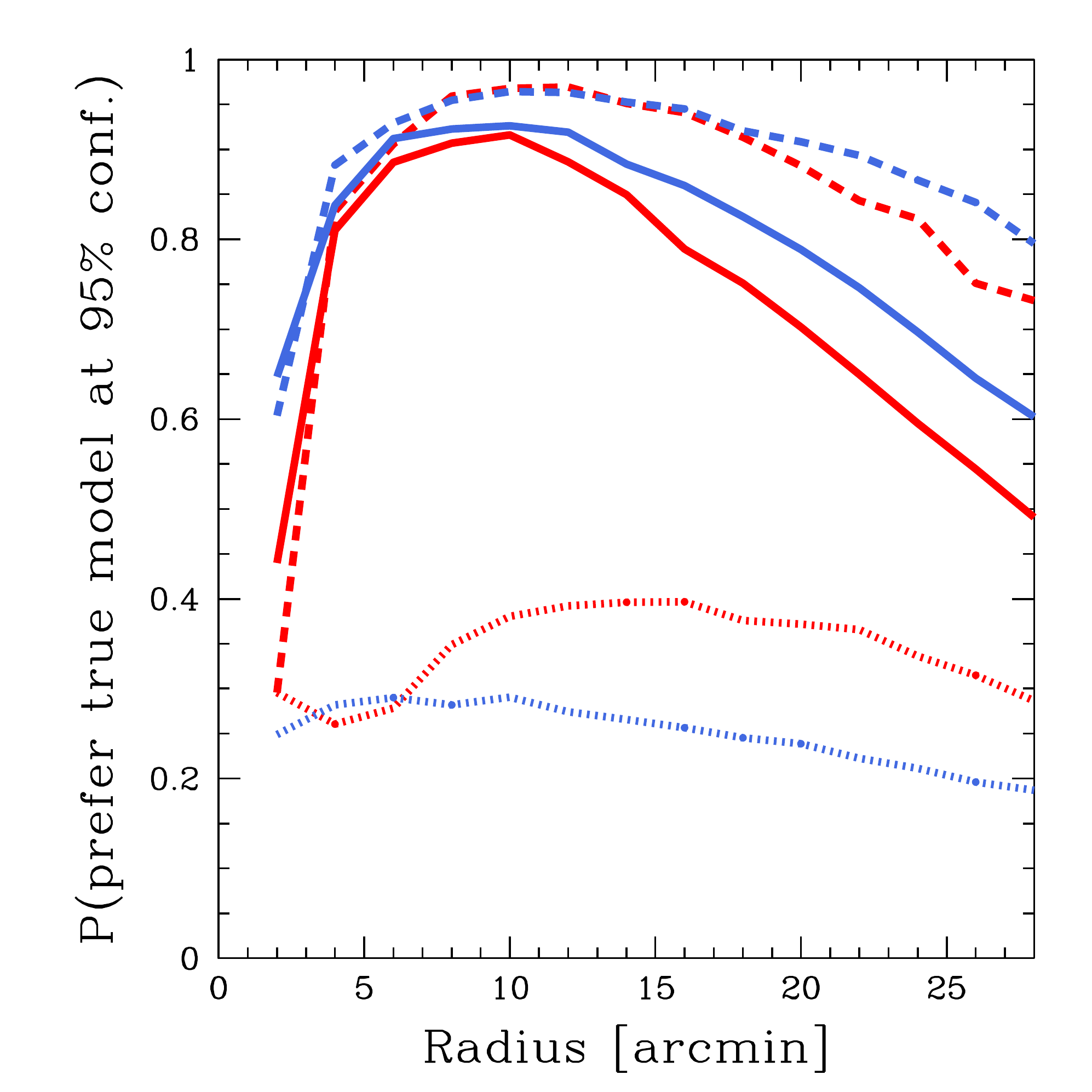}}\\
\end{center}
\caption{The curves show the probability of preferring the true model at $95\%$ confidence given a measurement of the number of $m_{NB816}<26$ LAEs in high $\teff$ fields as a function of the target area around the quasar sightline. The red and blue curves correspond to the $\GHI$ and $T_{\rm IGM}$ models as the ``truth," respectively. The dashed curves assume that the background density of LAEs is previously known to high precision, while the solid curves use LAEs in an annulus of $34 < \theta < 45$ arcmin to measure the background density (e.g. at the edge of an HSC field, ignoring vignetting for simplicity). The dotted curves correspond to the probability that either model is distinguishable from a random region of sky (i.e. without a background quasar sightline) at $95\%$ confidence.}
\label{fig:lae_prob}

\end{figure}

In Figure~\ref{fig:lae_prob} we show the resulting probabilities of a single observation of LAEs ($m_{NB816}<26$ in a field with $\teff>6.5$) favoring one model over the other at $95\%$ confidence as a function of target sky area. Red and blue curves correspond to the fluctuating $\GHI$ and $T_{\rm IGM}$ models as the ``truth," respectively. The dashed curves assume that we know the background density of LAEs perfectly, while the solid curves perform a differential measurement versus the outer annulus of a HSC field of view (i.e. using the CDFs of $\Sigma_{\rm LAE,target}/\Sigma_{\rm LAE,field}$). In general, a single measurement of LAEs within a radius of $6$--$12$ arcmin has a fairly high probability ($\sim90\%$) of preferring the true model over the alternative. The drop-off on scales $\ga10$ arcmin is due to the strong signal near the quasar line of sight being washed out by averaging with more normal regions farther from the line of sight -- we leave a more sophisticated statistical approach that uses the full radial number density profile to future work. The dotted curves show the probability that the region can be distinguished from a random patch of sky (with the same area) at $95\%$ confidence. This probability is significantly lower, only $\sim30$--$40\%$. This relatively low probability is due to the large cosmic variance of LAEs in our model, in qualitative agreement with the variations between different fields observed with Subaru/Suprime-Cam by \citet{Ouchi08}, which limits the statistical significance of any excess or deficit of LAEs relative to the field.

\section{Discussion \& Conclusion}

The $\sim110$ Mpc$/h$ GP trough observed by B15 requires a strong, large-scale difference in the temperature of  low-density gas or the strength of the ionizing background. While large-scale overdense regions are likely to be cooler than average, because they were reionized early \citep{D'Aloisio15}, large-scale voids are likely to have a weaker ionizing background (DF16). We have shown that determining whether giant GP troughs are overdense or underdense from large-scale \lya forest statistics alone is unlikely, but the corresponding large-scale density fluctuation should lead to a similarly large-scale excess or deficit of galaxies in the surrounding field. While a full spectroscopic inventory of bright LBGs within $\sim10$ arcmin should show a very strong signal, the outcome of less expensive observations of LBGs with photometric redshifts or narrowband-selected LAEs is more subtle. Nevertheless, because the two IGM models we consider make opposing predictions, it appears very likely that a single measurement of LAEs in the ULAS J0148+0600 field will be able to distinguish between the models.

In our predictions for LAEs we have neglected the potential impact of a fluctuating ionizing background on the escape of \lya photons from their local environments. As first shown by \citet{BH13}, and confirmed by others \citep{Mesinger15,Choudhury15,Kakiichi16}, self-shielded neutral gas in regions with low $\GHI$ should suppress the observability of LAEs. That is, the weak ionizing background required to produce a giant GP trough may act to suppress the number of LAEs even further than the large-scale void environment, although the strength of this effect depends on the model for escape of \lya emission from the galaxies themselves. If the fluctuating $\GHI$ model is correct, this effect should strengthen the signal (in LAEs) relative to our predictions. In contrast, $T_{\rm IGM}$ fluctuations are unlikely to have much of an effect because the thermal state of moderately-overdense gas capable of self-shielding is relatively insensitive to reionization heating, as shown by the curves in Figure~\ref{fig:tempevol} at $\Delta\ga10$.

While large-scale \lya forest transmission measurements are similar between the two IGM models, it is possible that smaller-scale transmission statistics and/or higher-order Lyman-series lines would be more constraining (e.g. \citealt{Davies17,Gnedin17}). Due to their smaller oscillator strengths, transmission in the higher-order Lyman-series forests is sensitive to gas at higher densities than the \lya forest \citep{OF05}, so they are less likely to be fully saturated (cf. the $\teff=5.17\pm0.05$ measured in the \lyb forest of the giant GP trough of ULAS J0148+0600). The two models also make predictions for the nature of regions with \emph{excess} transmission; in the fluctuating $\GHI$ model they are overdense, while in the fluctuating $T_{\rm IGM}$ model they are hot, underdense environments. The discovery of a sightline with extraordinarily \emph{low} $\teff$ at $z\sim5.7$, predicted to exist in the tail of both models we consider, would allow for a definitive test of our hypotheses, but no such regions have yet been identified. High-resolution spectroscopy of regions with strong transmission spikes may be able to distinguish between the two scenarios by measuring something akin to the curvature statistic \citep{Becker11a} or wavelet amplitudes \citep{LM14}. Another possible avenue of investigation is whether high-redshift low-ionization absorption systems \citep{Becker11b}, thought to be associated with low-mass galaxies \citep{Finlator13}, are correlated with $\teff$. However, the cross-section of such gas around halos is likely also dependent on the strength of the ionizing background and properties of galactic feedback \citep{Keating14,Keating16}, so any such correlations would be challenging to interpret.

Whichever model is correct (if either!) will have profound implications for our knowledge of the epoch of reionization. If post-reionization ionizing background fluctuations are strong, then models of the reionization process will have to contend with the effects of a short, spatially-varying mean free path which manifest on extremely large scales. If post-reionization temperature fluctuations are strong, then reionization must have been an extended process, and the required amount of heat injection places constraints on the spectra of sources of ionizing photons. Also important is that the alternative model is \emph{ruled out} due to the strength of the effect and their tendency to cancel each other out, i.e. either ionizing background or temperature fluctuations must be \emph{weak}, requiring a long mean free path of ionizing photons or a rapid/low heat input by reionization, respectively. In the future, one could imagine populating Figure~\ref{fig:taugal_bb} and/or Figure~\ref{fig:taugal_nb} with measurements of LBGs and/or LAEs in the foreground of several $z\ga6$ quasars, which should show either a positive or negative trend if either of the models we consider is valid. If no correlation between \lya opacity and galaxy density is seen, then other factors (e.g., \citealt{Chardin15}) may need to be considered.
 
\section*{Acknowledgements}

We would like to thank Anson D'Aloisio and Matthew McQuinn for comments on a draft of this manuscript. We would also like to thank Joseph Hennawi and the ENIGMA group at MPIA for helpful discussions.

GB acknowledges support from the National Science Foundation through grant AST-1615814. SRF was supported by NASA through award NNX15AK80G. Parts of this research were completed as part of the University of California Cosmic Dawn Initiative. S.R.F. acknowledge support from the University of California Office of the President Multicampus Research Programs and Initiatives through award MR-15-328388. 

\bibliographystyle{apj}
 \newcommand{\noop}[1]{}

\end{document}